\documentclass[12pt,a4paper]{article}
\pdfoutput=1
\usepackage[a4paper]{geometry}
\usepackage{graphicx}
\usepackage{amsmath}
\usepackage{amssymb}
\usepackage{float}
\usepackage{listings}
\usepackage{tikz}
\title{\bf Solution of the Bartels-Kwiecinski-Praszalowicz  equation via Monte Carlo integration}
\author{Grigorios Chachamis \&  Agust{\' \i}n Sabio Vera\\ \\
{\small Instituto de F{\' \i}sica Te{\' o}rica UAM/CSIC, Nicol{\'a}s Cabrera 15}\\ 
{\small \& Universidad Aut{\' o}noma de Madrid, E-28049 Madrid, Spain.}\\
}

\begin{document}

\maketitle 

\abstract
We present a method of solution of the Bartels-Kwiecinski-Praszalowicz (BKP) equation based on the numerical integration of iterated integrals in transverse momentum and rapidity space. As an application, our procedure, which makes use of Monte Carlo integration techniques, is applied to obtain the gluon Green function in the odderon case at leading order. The same approach can be used for more complicated scenarios. 

\section{Brief introduction to the BKP equation}

Two main actors rule the high energy behavior of scattering amplitudes in Quantum Chromodynamics: the pomeron and the odderon~\cite{Odderon}. The former dominates at asymptotic energies and the latter plays its role at lower energies. In perturbation theory both are described as compound states of reggeized gluons, two for the pomeron and three in the odderon case. These different structures are dictated by their corresponding behavior under charge conjugation: $C=+$ for pomeron and $C=-$ for odderon. The amount of work devoted to the study of the pomeron has been tantalizing. The investigation of the odderon is harder, both phenomenologically and theoretically. The experimental searches are complicated since the odderon generally corresponds to a subleading contribution to cross sections. From a theoretical point of view, it turns out that the integral equation describing the $t$-channel exchange of an odderon via the coupling of three off-shell reggeized gluons is rather complicated. However, it enjoys 2-dimensional conformal invariance in coordinate representation, which allows for the application of techniques previously developed for conformal field theory and integrable systems~\cite{Integrability}~\cite{Recent}. 

In the present work we propose to take a route quite different to previous approaches. We operate in transverse momentum and rapidity space and solve the equation governing the odderon Green function, the so-called Bartels-Kwiecinski-Praszalowicz  (BKP) equation~\cite{BKP}, by iterating it. The final outcome is a set of nested integrations which we evaluate using Monte Carlo integration techniques.  We focus in our work on the odderon case but we would like to emphasize that the techniques here described are readily applicable to those processes where a larger number of reggeized gluons is exchanged in the $t$-channel either at leading or higher orders. 

Let us write the BKP equation for three reggeons (the generalization to a larger number corresponds to a straightforward extension of this example) in the following form
\begin{eqnarray}
\left(\omega - \omega({\bf p}_1) - \omega({\bf p}_2) - \omega({\bf p}_3) \right) 
 f_\omega \left({\bf p}_1,{\bf p}_2,{\bf p}_3\right) &=& \nonumber\\
&& \hspace{-6cm} \delta^{(2)}  \left({\bf p}_1-{\bf p}_4 \right) \delta^{(2)}  \left({\bf p}_2-{\bf p}_5 \right) 
 \delta^{(2)}  \left({\bf p}_3-{\bf p}_6 \right)  \nonumber\\
 &&\hspace{-6cm}+  \int d^2 {\bf k} \, \xi \left({\bf p}_1,{\bf p}_2,{\bf p}_3,{\bf k}\right)
 f_\omega \left({\bf p}_1+{\bf k},{\bf p}_2-{\bf k},{\bf p}_3\right) \nonumber\\
 &&\hspace{-6cm}+  \int d^2 {\bf k} \,  \xi \left({\bf p}_2,{\bf p}_3,{\bf p}_1,{\bf k}\right)
  f_\omega \left({\bf p}_1,{\bf p}_2+{\bf k},{\bf p}_3-{\bf k} \right) \nonumber\\
 &&\hspace{-6cm}+  \int d^2 {\bf k}  \, \xi \left({\bf p}_1,{\bf p}_3,{\bf p}_2,{\bf k}\right)f_\omega \left({\bf p}_1+{\bf k},{\bf p}_2,{\bf p}_3-{\bf k}\right),
\label{BKPeq} 
\end{eqnarray}
with the strong coupling being ${\bar \alpha}_s = \alpha_s N_c / \pi$, the square of a Lipatov's emission vertex
\begin{eqnarray}
\xi \left({\bf p}_1,{\bf p}_2,{\bf p}_3,{\bf k}\right) ~= \frac{{\bar \alpha}_s}{4} \frac{ \theta({\bf k}^2 - \lambda^2)}{\pi {\bf k}^2}
 \Bigg(1+\frac{({\bf p}_1+{\bf k})^2 {\bf p}_2^2-({\bf p}_1+{\bf p}_2)^2 {\bf k}^2}{{\bf p}_1^2 ({\bf k}-{\bf p}_2)^2}\Bigg),
 \label{xifunction}
\end{eqnarray}
and the gluon Regge trajectory at leading order 
\begin{eqnarray}
\omega ({\bf p}) &=& - \frac{{\bar \alpha}_s}{2} \ln{\frac{{\bf p}^2}{\lambda^2}}.
\end{eqnarray}
We use the shorter notation $f_\omega \left({\bf p}_1,{\bf p}_2,{\bf p}_3 \right)$ for 
$f_\omega \left({\bf p}_1,{\bf p}_2,{\bf p}_3;{\bf p}_4,{\bf p}_5,{\bf p}_6\right)$ where ${\bf p}_i$ are two-dimensional transverse vectors. This Green function $f_\omega$ describes the transition from three off-shell gluons with momenta ${\bf p}_{i=1,2,3}$ and rapidity $Y$ to three with momenta ${\bf p}_{i=4,5,6}$ and rapidity 0 (in our normalization). The dependence on $Y$ is hidden in the variable $\omega$ which we will remove later. $\lambda$ is a regulator of infrared divergencies whose influence is negligible as we take $\lambda \to 0$. 

\begin{figure}[h]
\begin{center}
\includegraphics[width=4.cm]{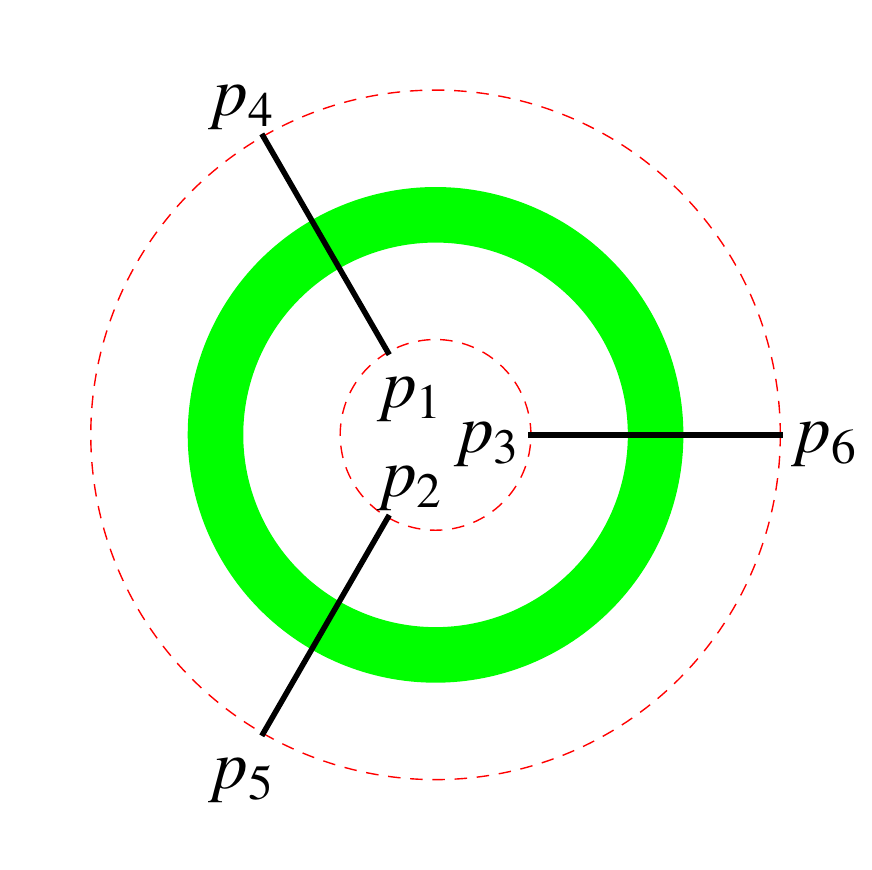}\includegraphics[width=.5cm]{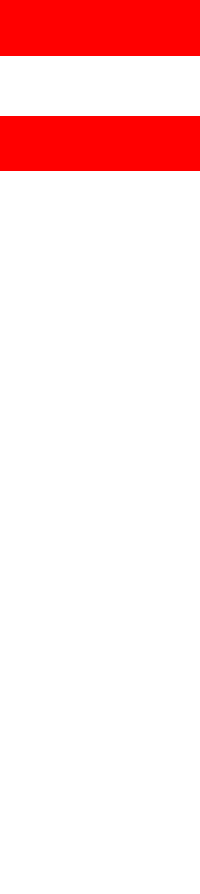}\includegraphics[width=4.cm]{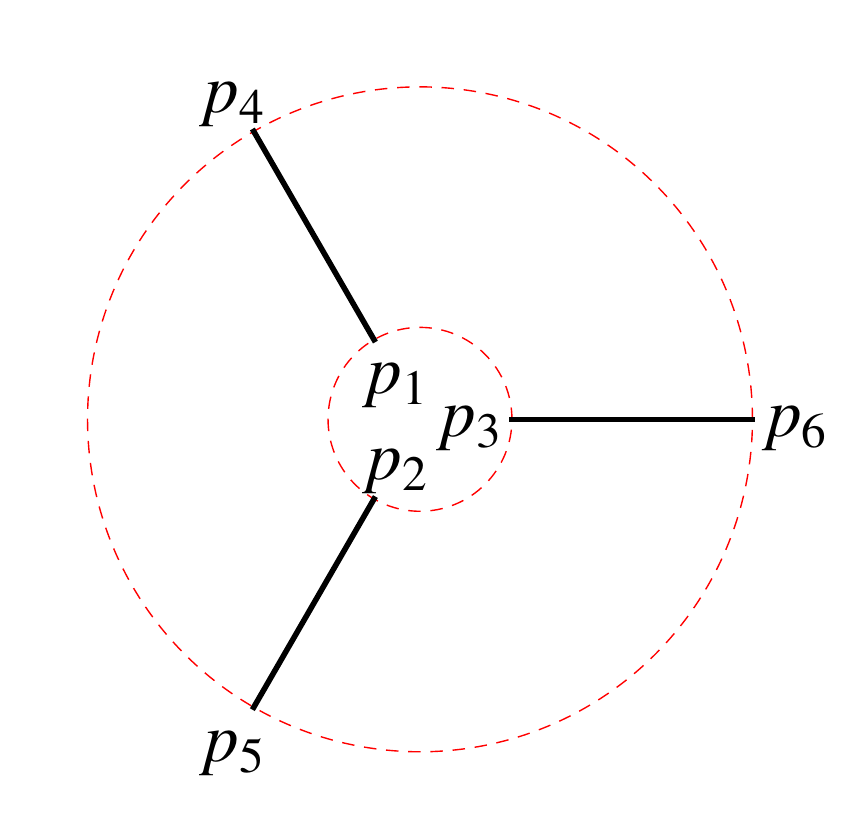}
\hspace{3cm}
\includegraphics[width=.5cm]{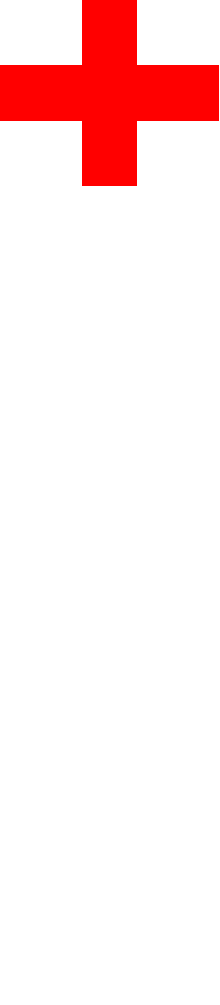}\includegraphics[width=4cm]{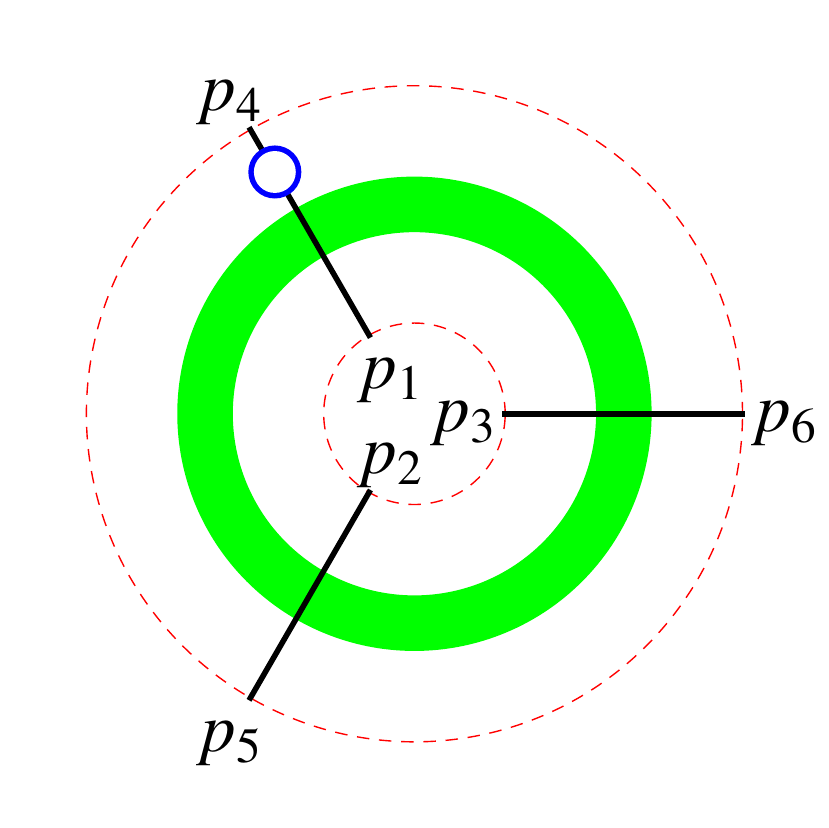}\includegraphics[width=.5cm]{Plus.pdf}\includegraphics[width=4cm]{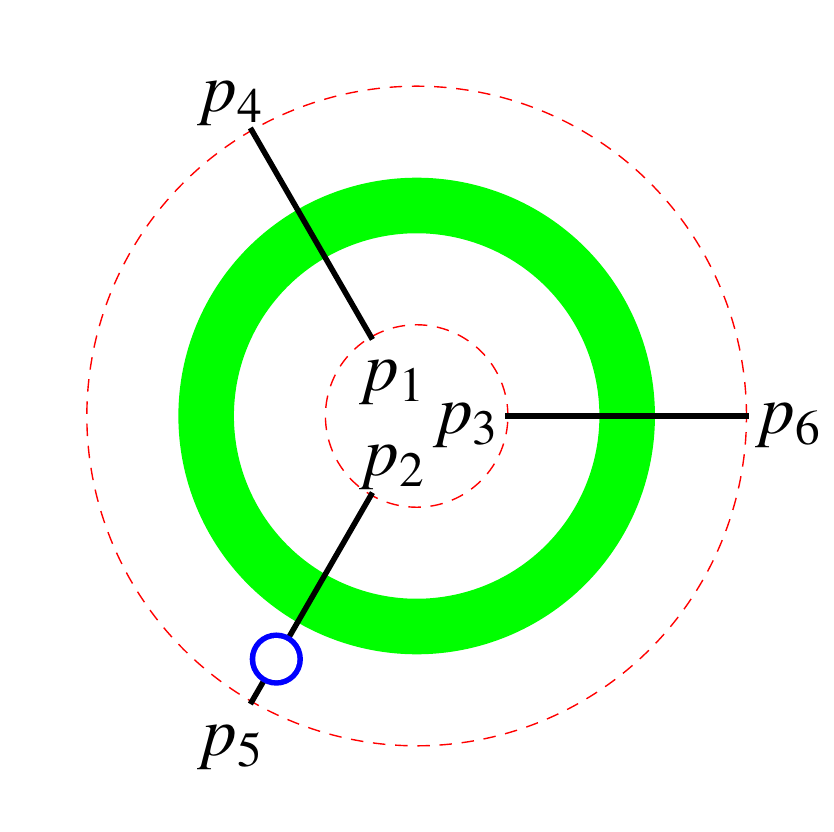}\includegraphics[width=.5cm]{Plus.pdf}\includegraphics[width=4cm]{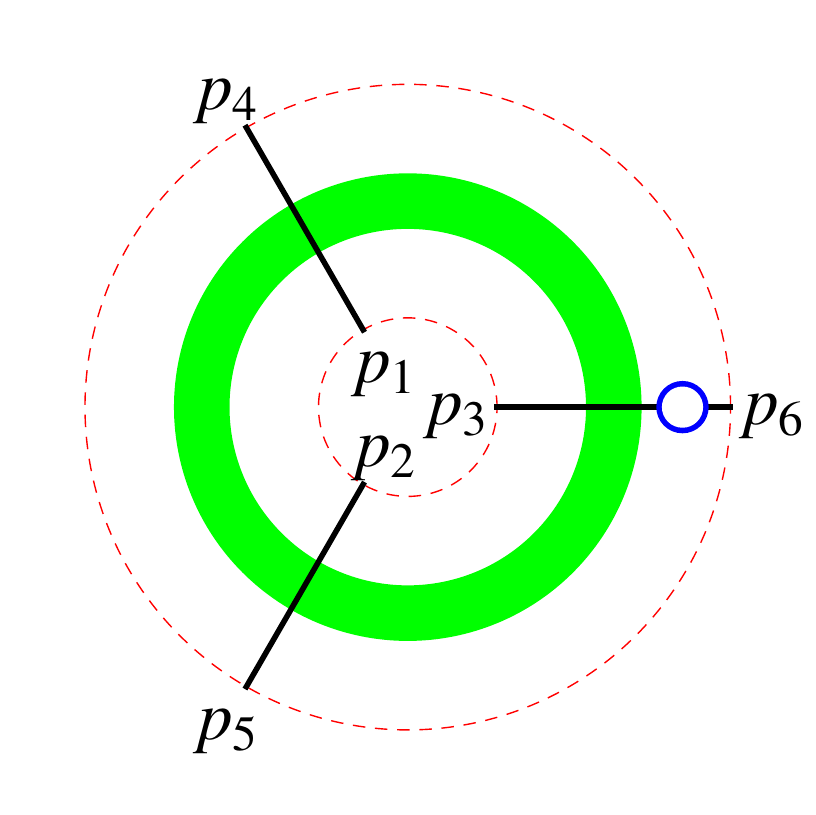}\hspace{3cm}
\includegraphics[width=.5cm]{Plus.pdf}\includegraphics[width=4cm]{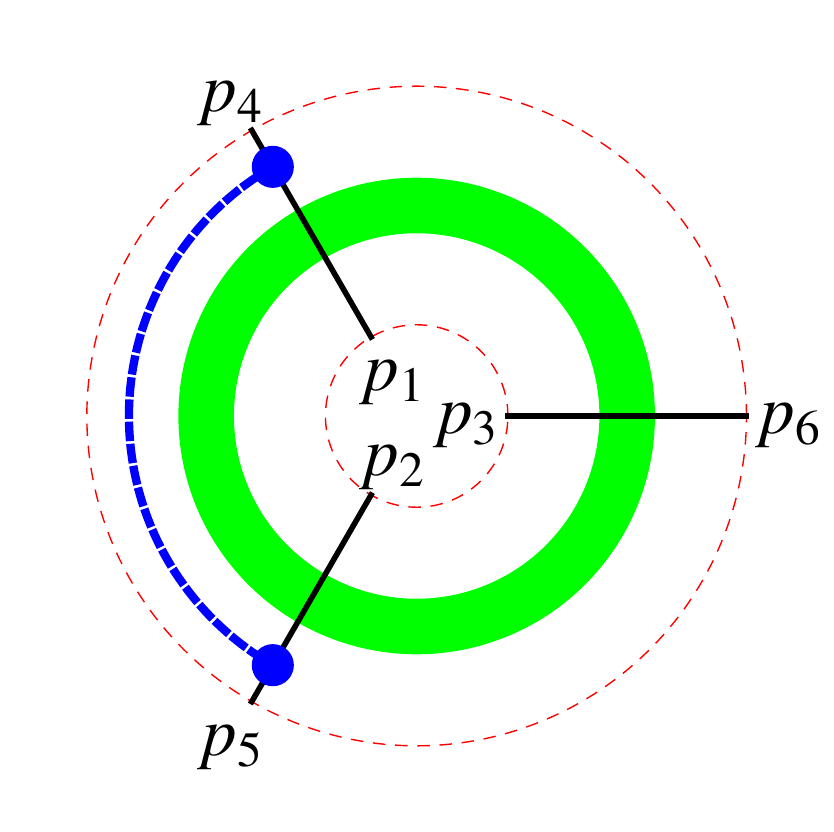}\includegraphics[width=.5cm]{Plus.pdf}\includegraphics[width=4cm]{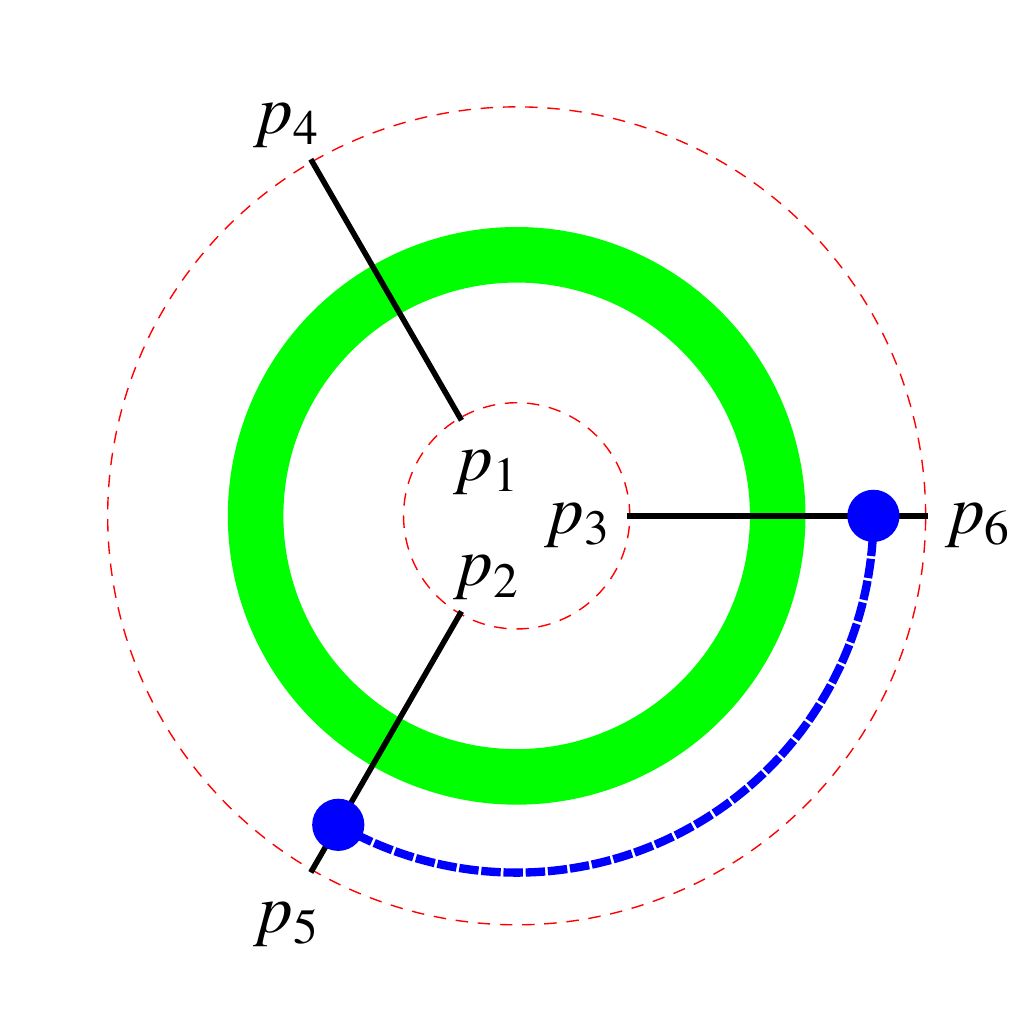}\includegraphics[width=.5cm]{Plus.pdf}\includegraphics[width=4cm]{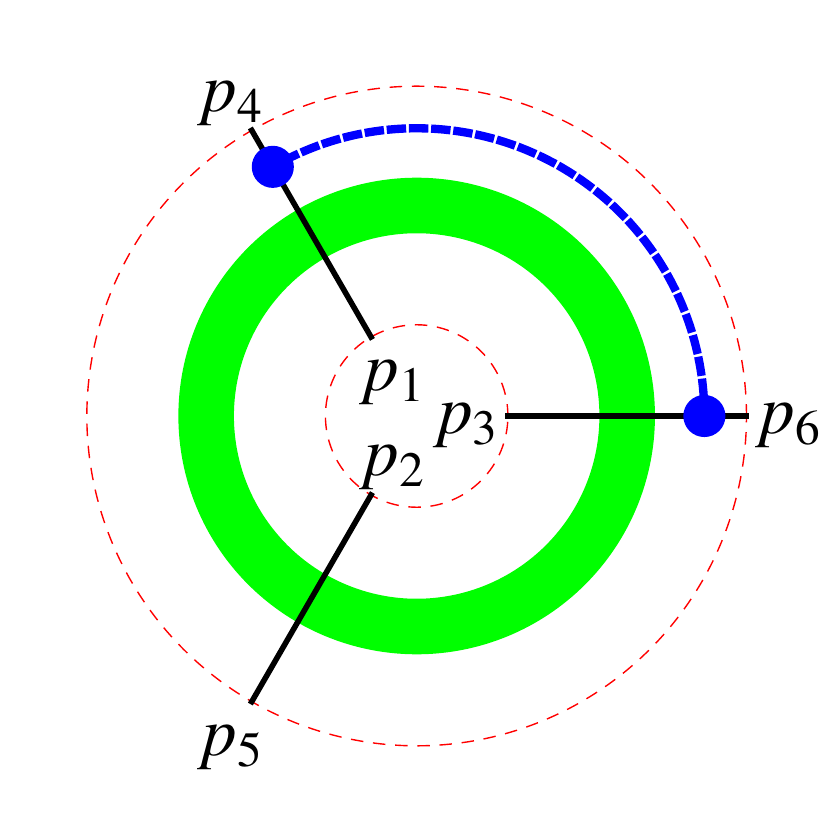}
\caption{BKP equation for the gluon Green function $f_\omega \left({\bf p}_1,{\bf p}_2,{\bf p}_3;{\bf p}_4,{\bf p}_5,{\bf p}_6\right)$ as in Eq.~(\ref{BKPeq}).}
\end{center}
\label{BKPeqGraph}
\end{figure}

 A graphical representation of this equation is shown in Fig.~\ref{BKPeqGraph}. 
The inner dashed circle corresponds to particles with a rapidity $Y$ and the outer one to those with rapidity $0$. Let us remark that the function $\xi \left({\bf p}_i,{\bf p}_j,{\bf p}_k,{\bf k}\right) $ corresponds to the coupling of two reggeized gluons with momenta ${\bf p}_i$ and ${\bf p}_j$ via a normal gluon with transverse momentum ${\bf k}$, leaving the third reggeized gluon, with momentum ${\bf p}_k$, as a simple spectator. In the LHS of Eq.~(\ref{BKPeq}) there are three terms proportional to the trajectory $\omega ({\bf p}_i)$ which are represented by the three contributions in the second line of Fig.~\ref{BKPeqGraph}. The function $\xi$ appears in the last three terms of the same figure. The first term in the RHS of Eq.~(\ref{BKPeq}) represents the initial condition for evolution in rapidity with the three two-dimensional Dirac delta functions $\delta^{(2)}  \left({\bf p}_1-{\bf p}_4 \right) \delta^{(2)}  \left({\bf p}_2-{\bf p}_5 \right)  \delta^{(2)}  \left({\bf p}_3-{\bf p}_6 \right)$.

We will now explain in some detail our procedure to solve Eq.~(\ref{BKPeq}). In a nutshell, we first iterate it in $\omega$ representation to then transform the result to get back to a representation with only transverse momenta and rapidity. 

\section{Solution in transverse momentum \& rapidity space}

In order to describe our method of solution we first streamline the notation using the operator
\begin{eqnarray}
{\cal O} ({\bf k}) \otimes f \left({\bf p}_1,{\bf p}_2,{\bf p}_3\right) & \equiv& \xi \left({\bf p}_1,{\bf p}_2,{\bf p}_3,{\bf k}\right)
 f \left({\bf p}_1+{\bf k},{\bf p}_2-{\bf k},{\bf p}_3\right) \nonumber\\
 &+& \xi \left({\bf p}_2,{\bf p}_3,{\bf p}_1,{\bf k}\right)
  f \left({\bf p}_1,{\bf p}_2+{\bf k},{\bf p}_3-{\bf k} \right) \nonumber\\
 &+&  \xi \left({\bf p}_1,{\bf p}_3,{\bf p}_2,{\bf k}\right)f \left({\bf p}_1+{\bf k},{\bf p}_2,{\bf p}_3-{\bf k}\right),
 \label{eq:bkpkernel} 
\end{eqnarray}
which allows us to write the BKP equation as
\begin{eqnarray}
\left(\omega - \omega({\bf p}_1) - \omega({\bf p}_2) - \omega({\bf p}_3) \right) 
 f_\omega \left({\bf p}_1,{\bf p}_2,{\bf p}_3\right) &=& \nonumber\\
&& \hspace{-8.cm} \delta^{(2)}  \left({\bf p}_1-{\bf p}_4 \right) \delta^{(2)}  \left({\bf p}_2-{\bf p}_5 \right) 
 \delta^{(2)}  \left({\bf p}_3-{\bf p}_6 \right) \nonumber\\
&& \hspace{-7.cm} +\int d^2 {\bf k} \,  {\cal O} ({\bf k}) \otimes f_\omega \left({\bf p}_1,{\bf p}_2,{\bf p}_3\right).
\end{eqnarray}
Moving the terms depending on $\omega$ and the gluon Regge trajectories to the denominator of the RHS and iterating the action of the integral operator, we can present the gluon Green function as a sum over the number of rungs, $n$, joining two reggeized gluons, each of them carrying a function $\xi$:
\begin{eqnarray}
 f_\omega \left({\bf p}_1,{\bf p}_2,{\bf p}_3\right) &=&   \frac{ \left(1+\sum_{n=1}^\infty \prod_{i=1}^n \int d^2 {\bf k}_i {\cal O} ({\bf k}_i) \otimes \right)  }{\left(\omega - \omega({\bf p}_1) - \omega({\bf p}_2) - \omega({\bf p}_3) \right) } \nonumber\\
 &&\delta^{(2)}  \left({\bf p}_1-{\bf p}_4 \right) \delta^{(2)}  \left({\bf p}_2-{\bf p}_5 \right) \delta^{(2)}  \left({\bf p}_3-{\bf p}_6 \right).
\end{eqnarray}
In order to operate in rapidity space we use the Mellin transform
\begin{eqnarray}
f \left({\bf p}_1,{\bf p}_2,{\bf p}_3, Y\right) &=& \int_{a-i \infty}^{a + i \infty} \frac{d \omega}{2 \pi i}  e^{\omega Y} 
f_\omega \left({\bf p}_1,{\bf p}_2,{\bf p}_3\right)  
\end{eqnarray}
and the relation for multiple poles
\begin{eqnarray}
\int_{a-i \infty}^{a + i \infty} \frac{d \omega}{2 \pi i} e^{\omega Y} \prod_{i=0}^n \frac{1}{(\omega-\omega_i)} &=&
e^{\omega_0 Y} \prod_{i=1}^n \int_0^{y_{i-1}} d y_i e^{(\omega_i - \omega_{i-1})y_i},
\end{eqnarray}
with $y_0 = Y$ and $n>0$, to express the gluon Green function as a sum over nested integrals in rapidity and integrals over two-dimensional transverse momenta, {\it i.e.},
\begin{eqnarray}
f \left({\bf p}_1,{\bf p}_2,{\bf p}_3, Y\right)  &=&   \nonumber\\
&&\hspace{-3cm}e^{(\omega({\bf p}_1) + \omega({\bf p}_2) + \omega({\bf p}_3)) Y} \delta^{(2)}  \left({\bf p}_1-{\bf p}_4 \right) \delta^{(2)}  \left({\bf p}_2-{\bf p}_5 \right) \delta^{(2)}  \left({\bf p}_3-{\bf p}_6 \right)
\nonumber\\
 &&\hspace{-3cm} + \sum_{n=1}^\infty \Bigg\{\prod_{i=1}^n \int_0^{y_{i-1}} d y_i 
 \int d^2 {\bf k}_i e^{(\omega({\bf p}_1) + \omega({\bf p}_2) + \omega({\bf p}_3)) (y_{i-1}- y_i)} {\cal O} ({\bf k}_i) \otimes \Bigg\} \nonumber\\
 &&\hspace{-2.5cm}e^{(\omega({\bf p}_1) + \omega({\bf p}_2) + \omega({\bf p}_3)) y_n} 
 \delta^{(2)}  \left({\bf p}_1-{\bf p}_4 \right) \delta^{(2)}  \left({\bf p}_2-{\bf p}_5 \right) \delta^{(2)}  \left({\bf p}_3-{\bf p}_6 \right).
 \label{IterativeEqn}
\end{eqnarray}

\begin{figure}[h]
\begin{center}
\includegraphics[width=6cm]{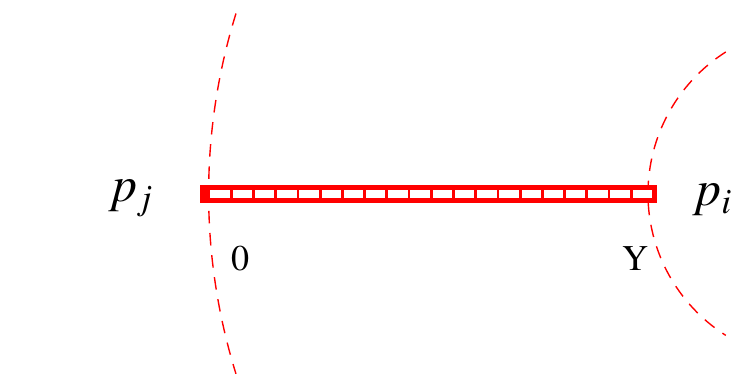} \, \, \, \includegraphics[width=.4cm]{Equal.pdf}\includegraphics[width=6cm]{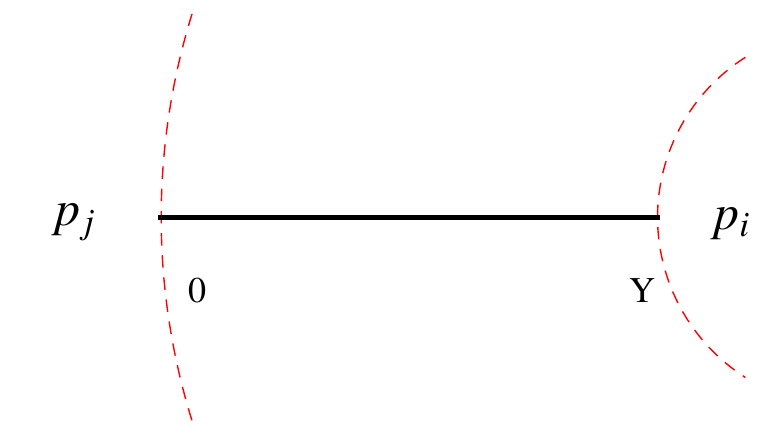}
\hspace{3cm}
\includegraphics[width=.4cm]{Plus.pdf}\includegraphics[width=6cm]{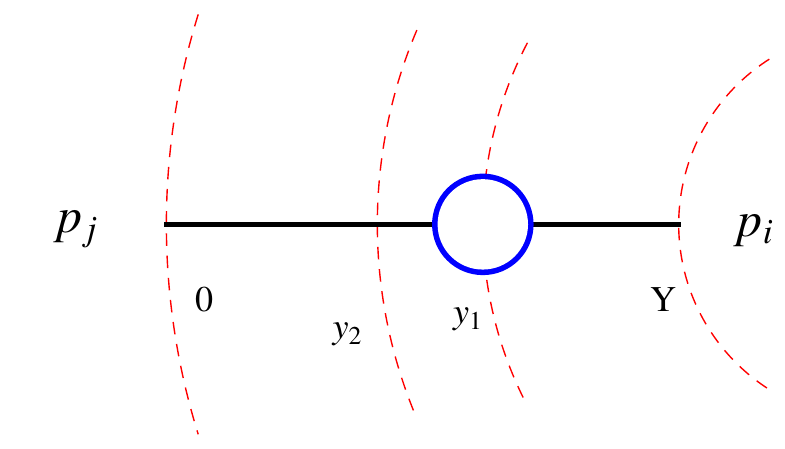} \, \, \,\includegraphics[width=.4cm]{Plus.pdf}\includegraphics[width=6cm]{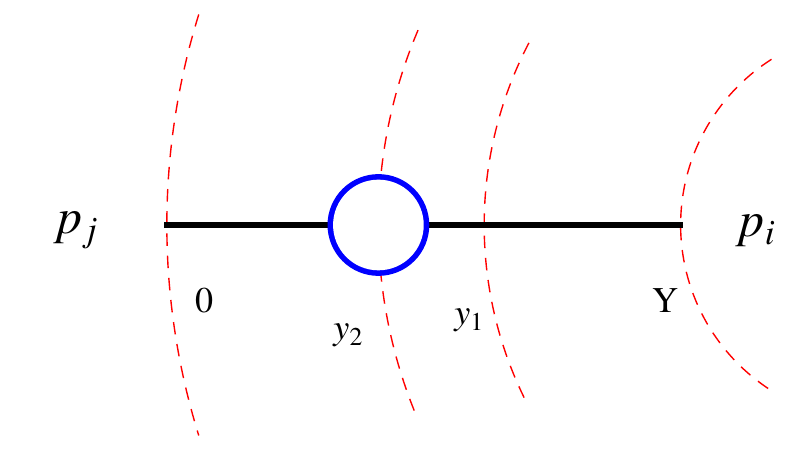}
\hspace{3cm}
\includegraphics[width=.4cm]{Plus.pdf}\includegraphics[width=6cm]{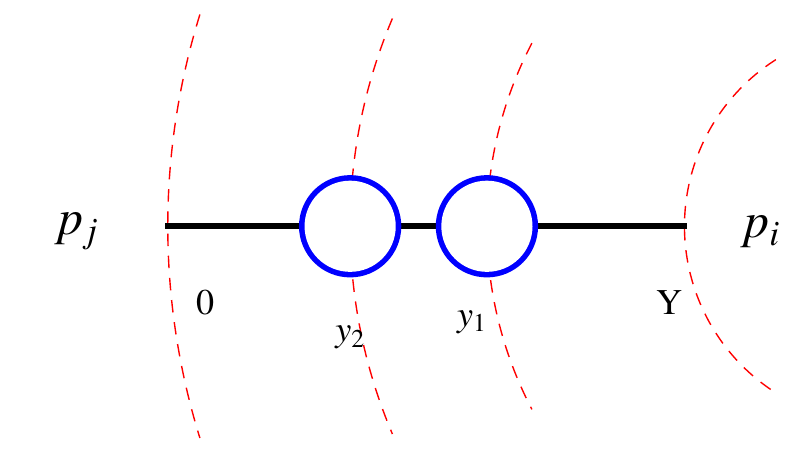} \, \, \,\includegraphics[width=.4cm]{Plus.pdf} \, \, \, \includegraphics[width=.54cm]{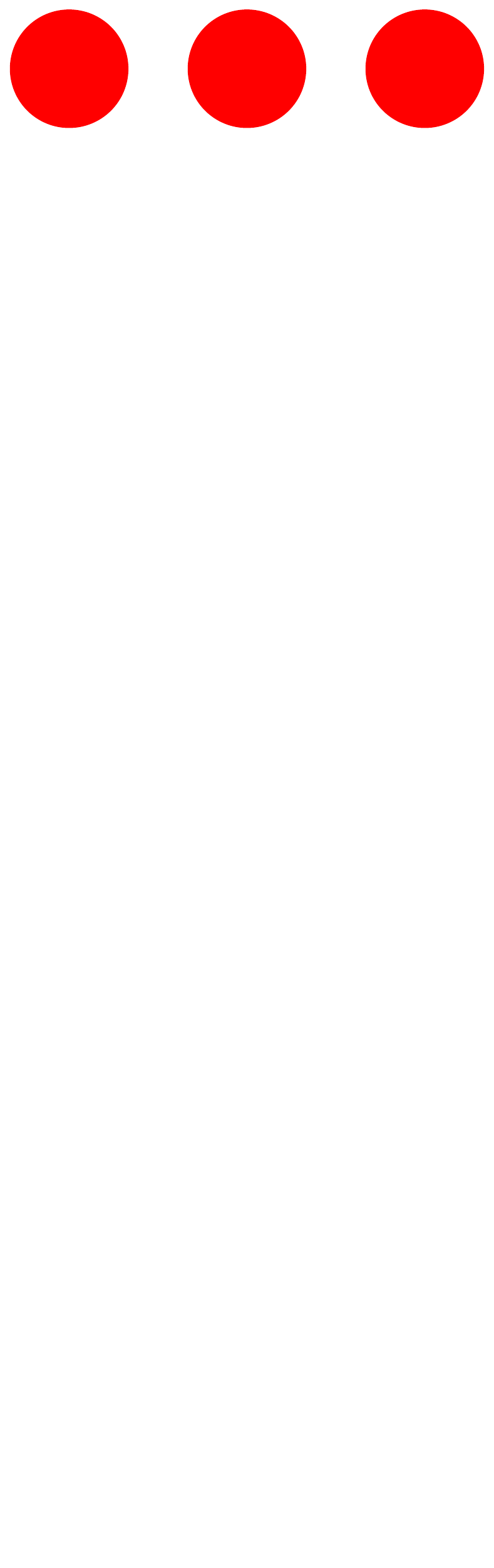} 
\end{center}
\caption{Reggeized gluon propagator containing ordered--in--rapidity Regge trajectories ($Y > y_{i} > y_{i+1} > 0$). }
\label{ReggPropas1}
\end{figure}

It is now mandatory to explain an important point. This solution is $\lambda$--independent, in the $\lambda \to 0$ limit, term--by--term in a perturbative expansion in ${\bar \alpha}_s$ (in the leading logarithmic approximation it is really an expansion in ${\bar \alpha}_s Y$).  However, this expansion is not convenient if our target is to obtain the full Green function since it generates a huge number of terms at each order in the coupling. The problem simplifies greatly if we resum the diagrams with gluon Regge trajectories using the notation explained in Fig.~\ref{ReggPropas1}. 

\begin{figure}
\begin{center}
\includegraphics[width=4.cm]{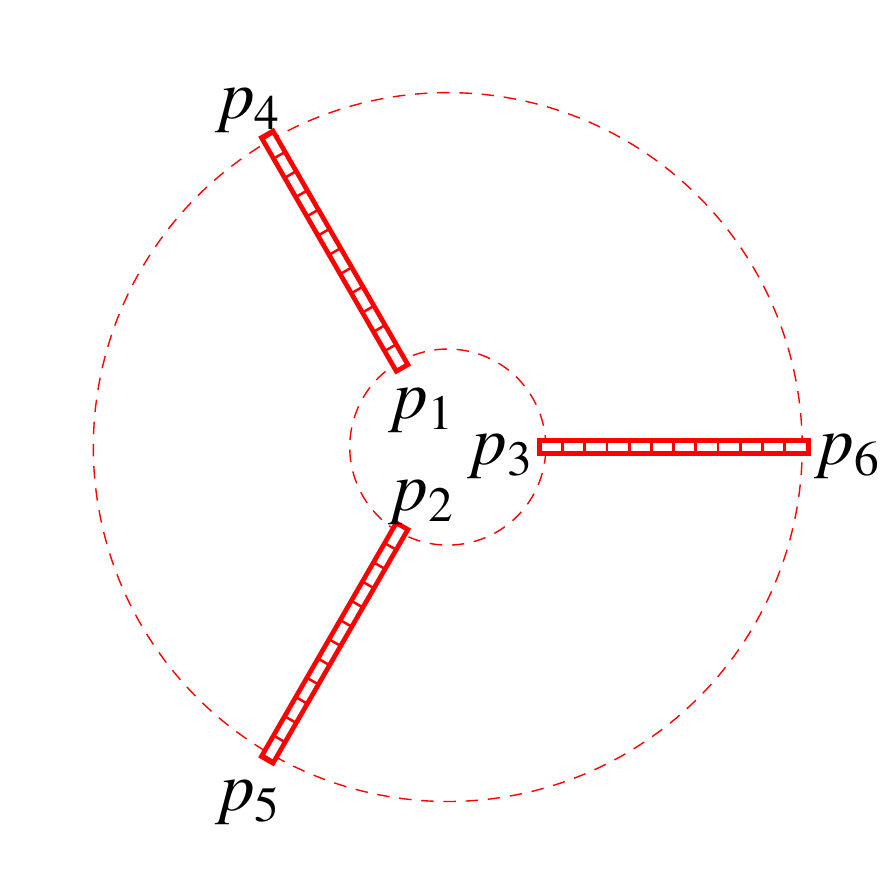}
\hspace{3cm}
\includegraphics[width=4cm]{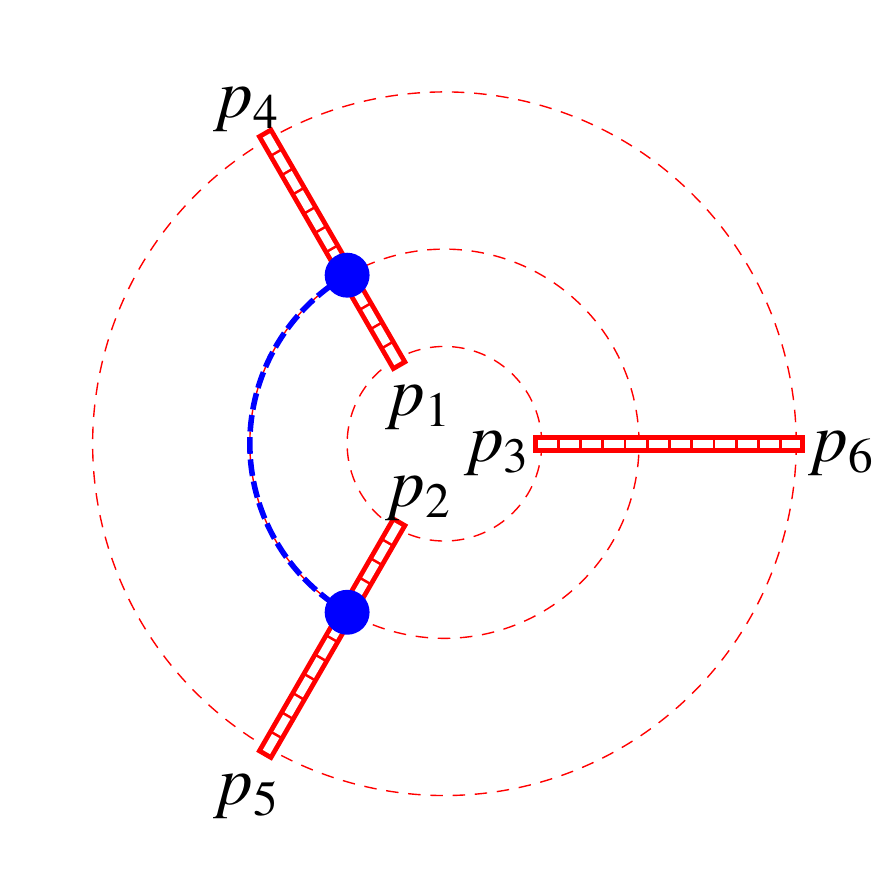}\includegraphics[width=4cm]{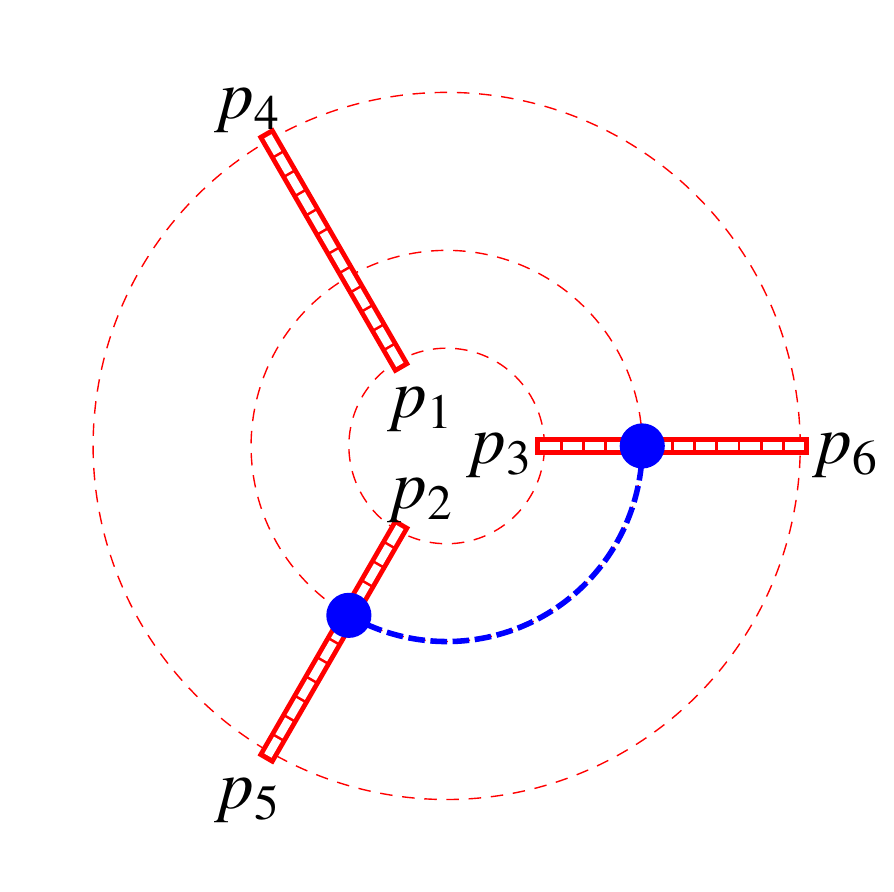}\includegraphics[width=4cm]{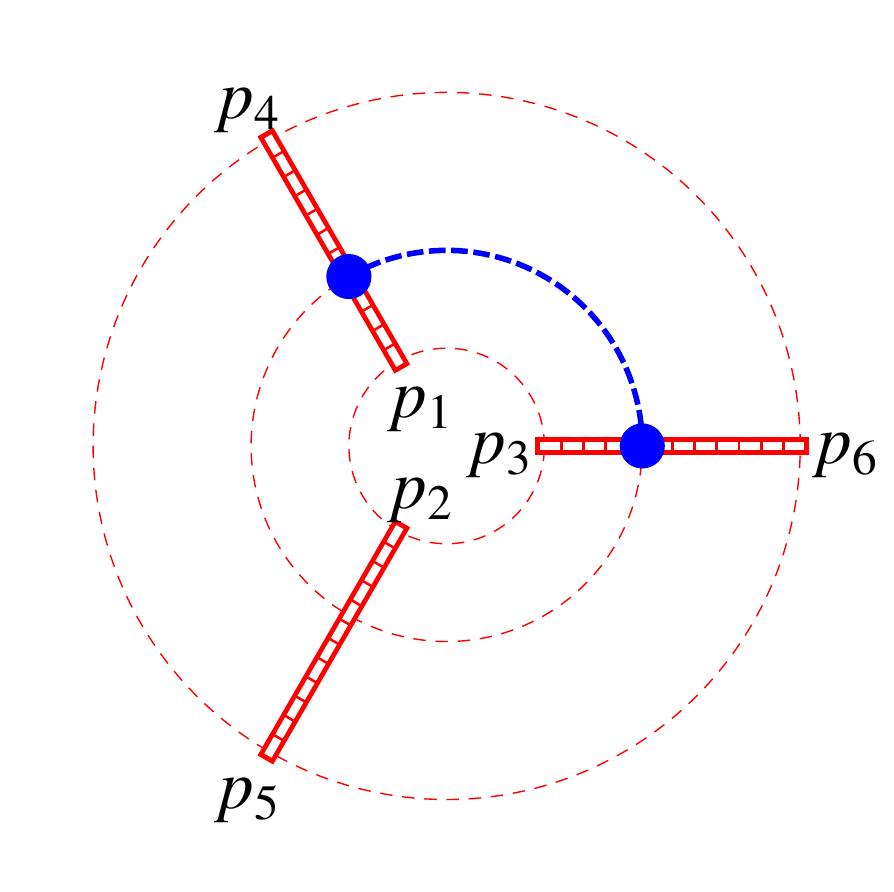}
\hspace{3cm}
\includegraphics[width=4cm]{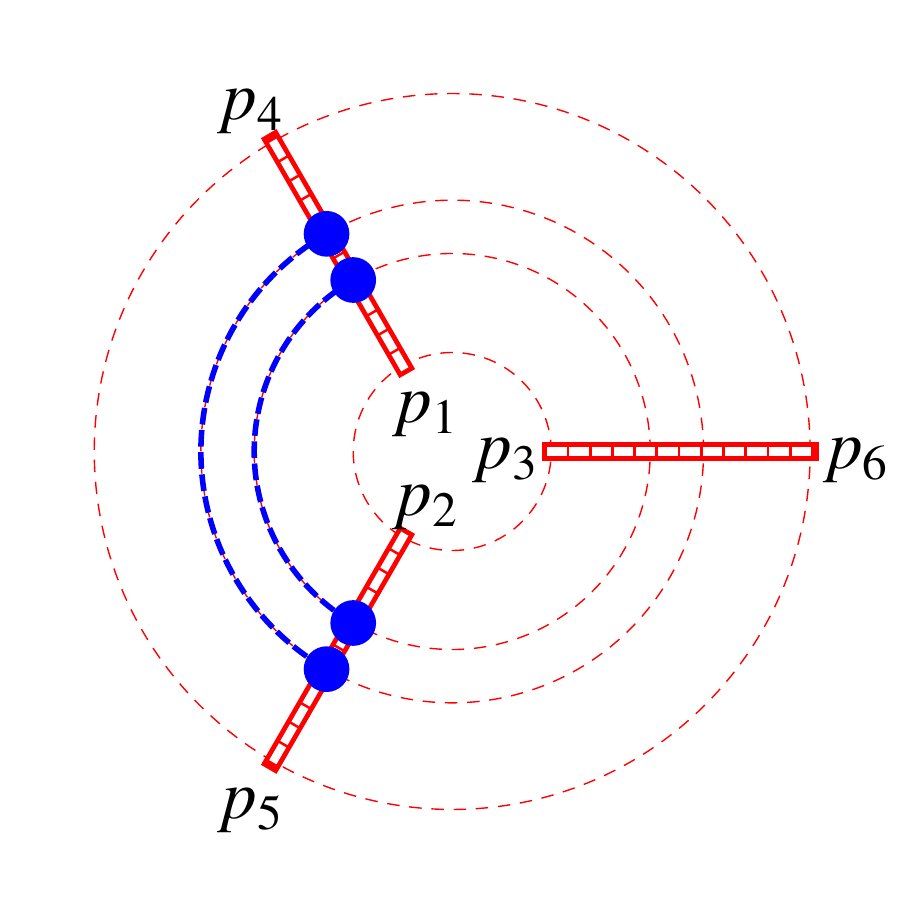}\includegraphics[width=4cm]{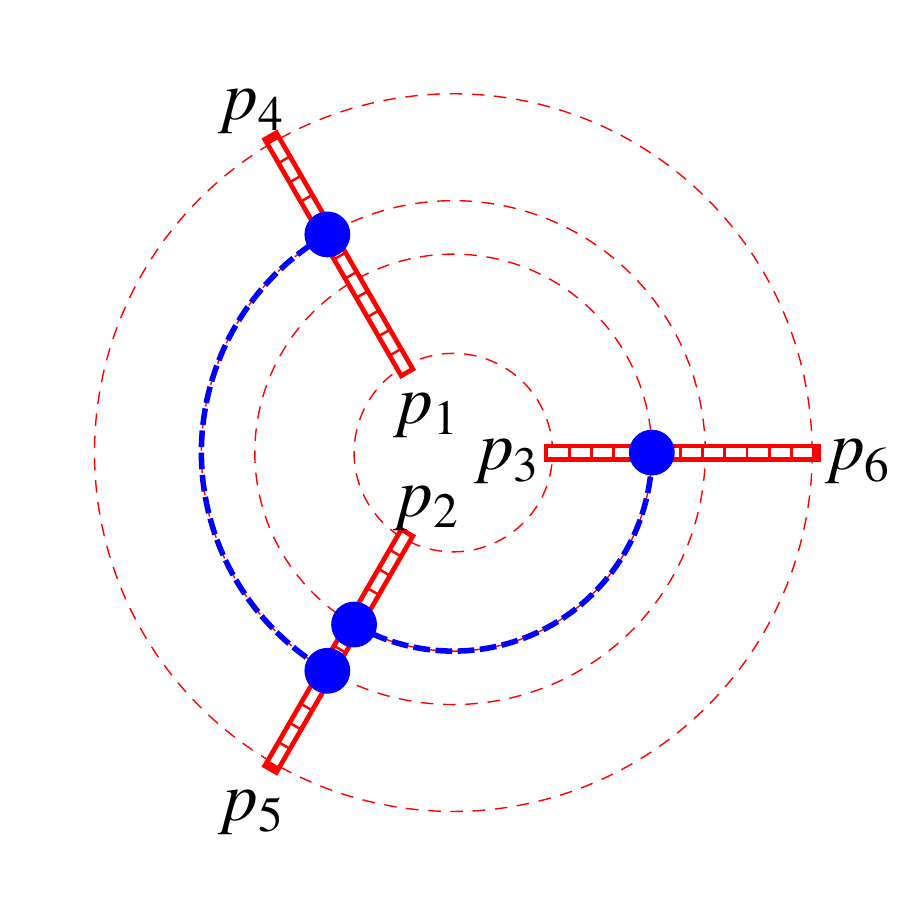}\includegraphics[width=4cm]{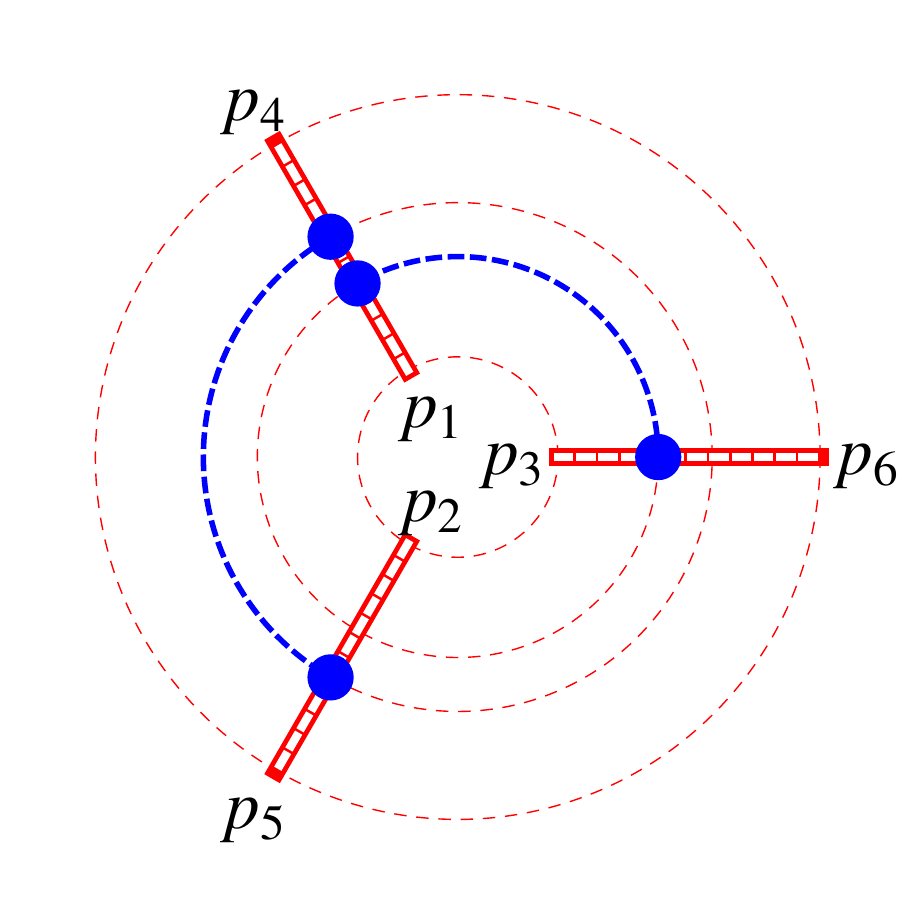}
\hspace{3cm}
\includegraphics[width=4cm]{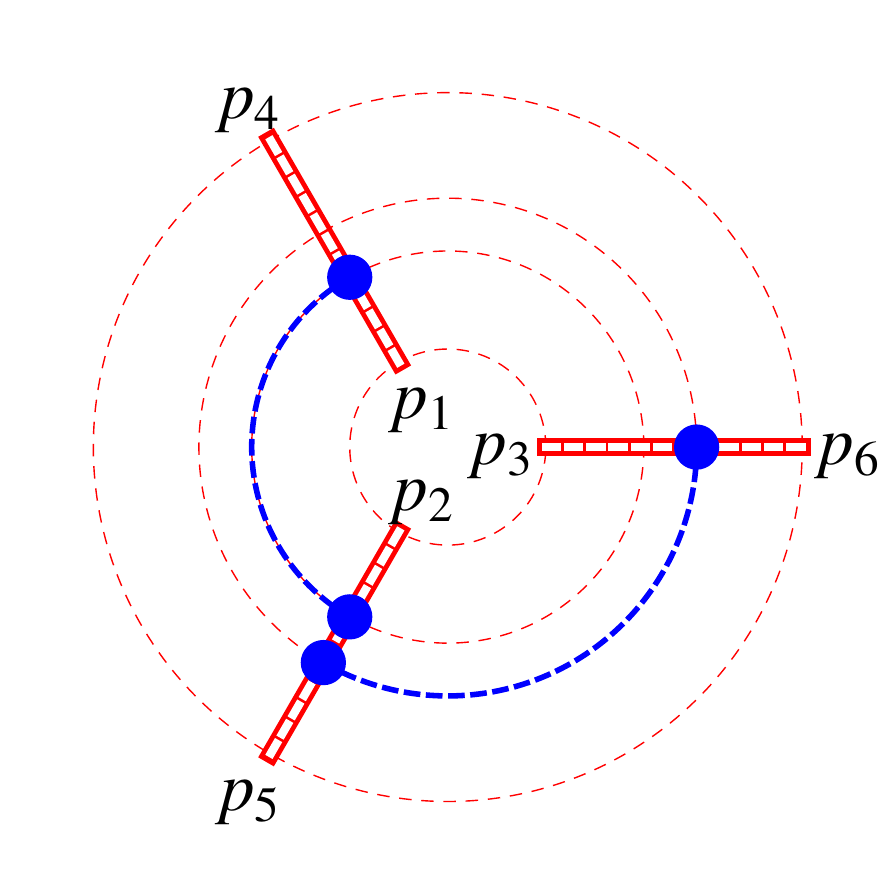}\includegraphics[width=4cm]{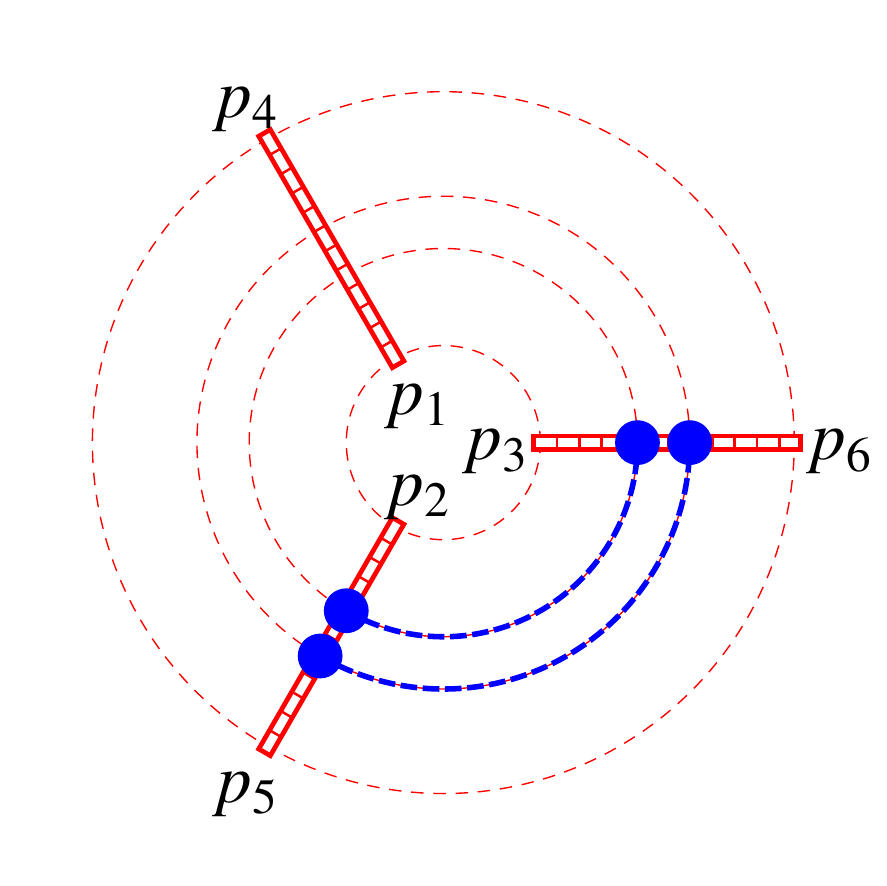}\includegraphics[width=4cm]{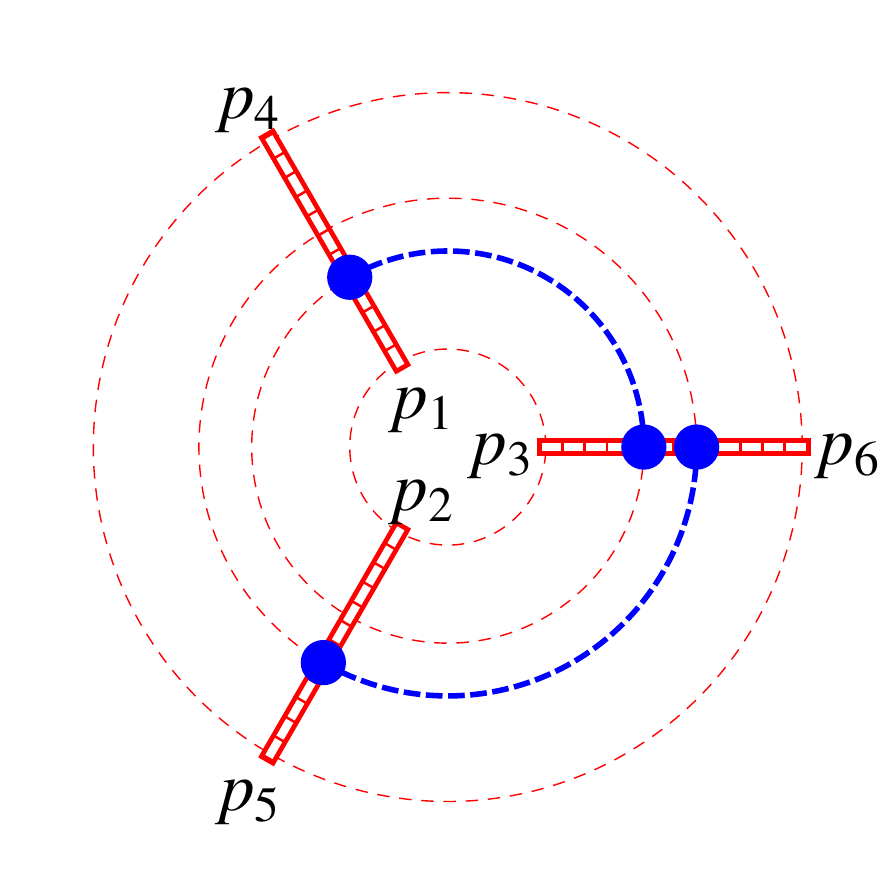}
\hspace{3cm}
\includegraphics[width=4cm]{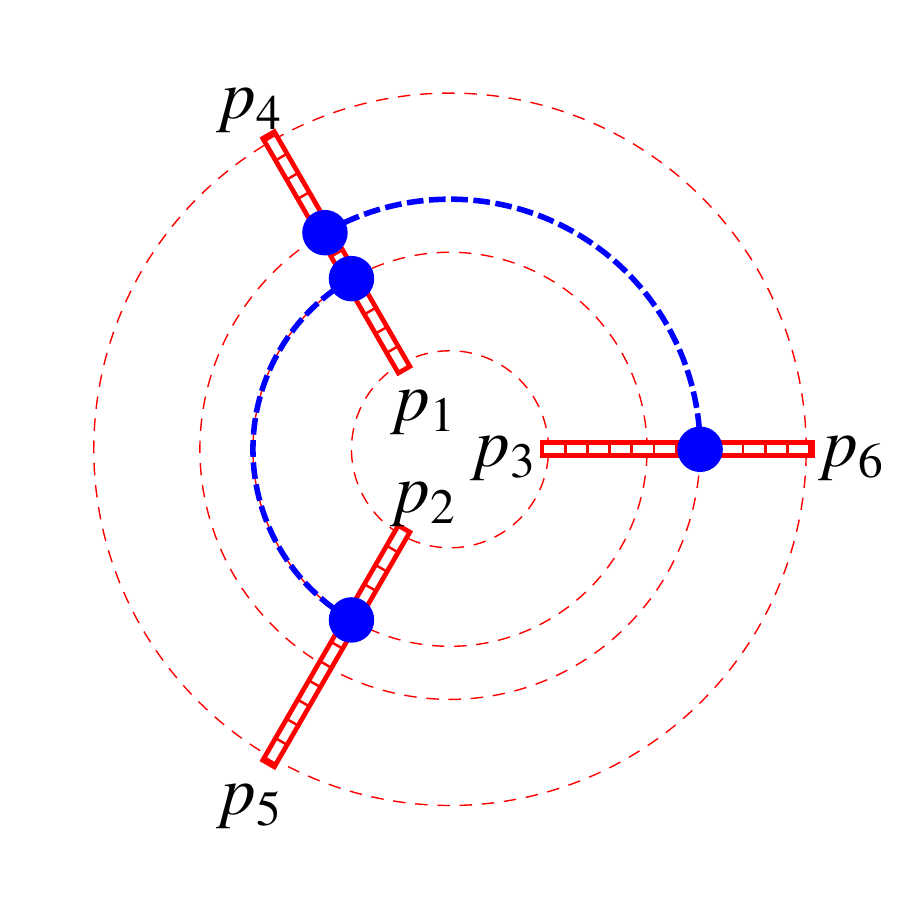}\includegraphics[width=4cm]{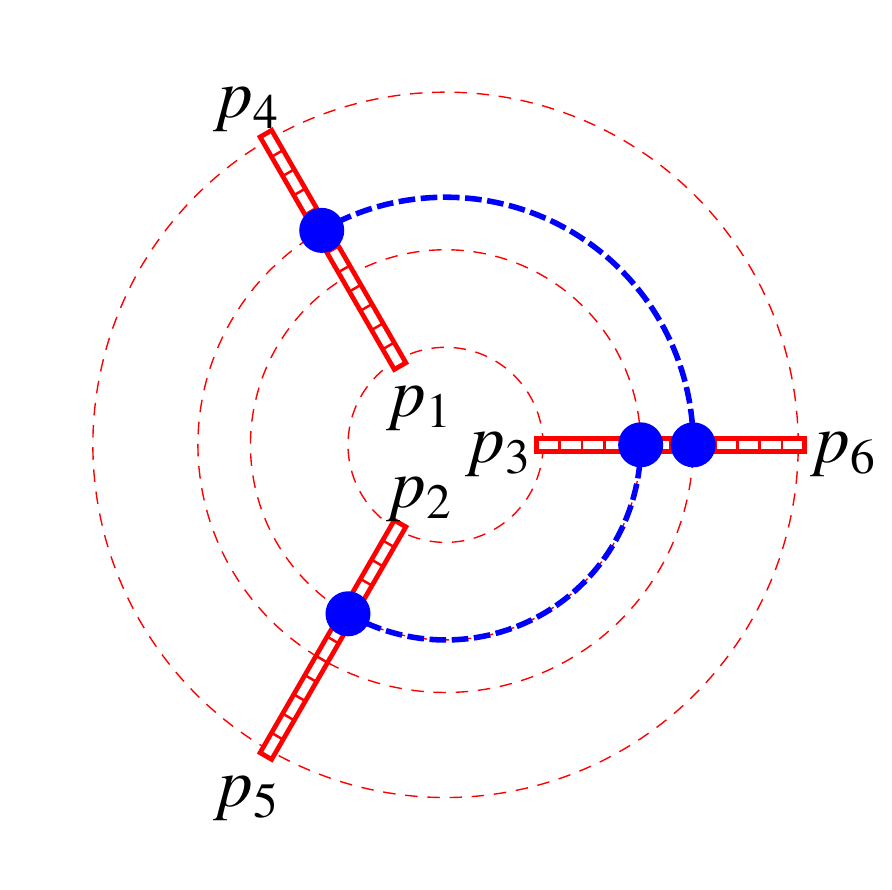}\includegraphics[width=4cm]{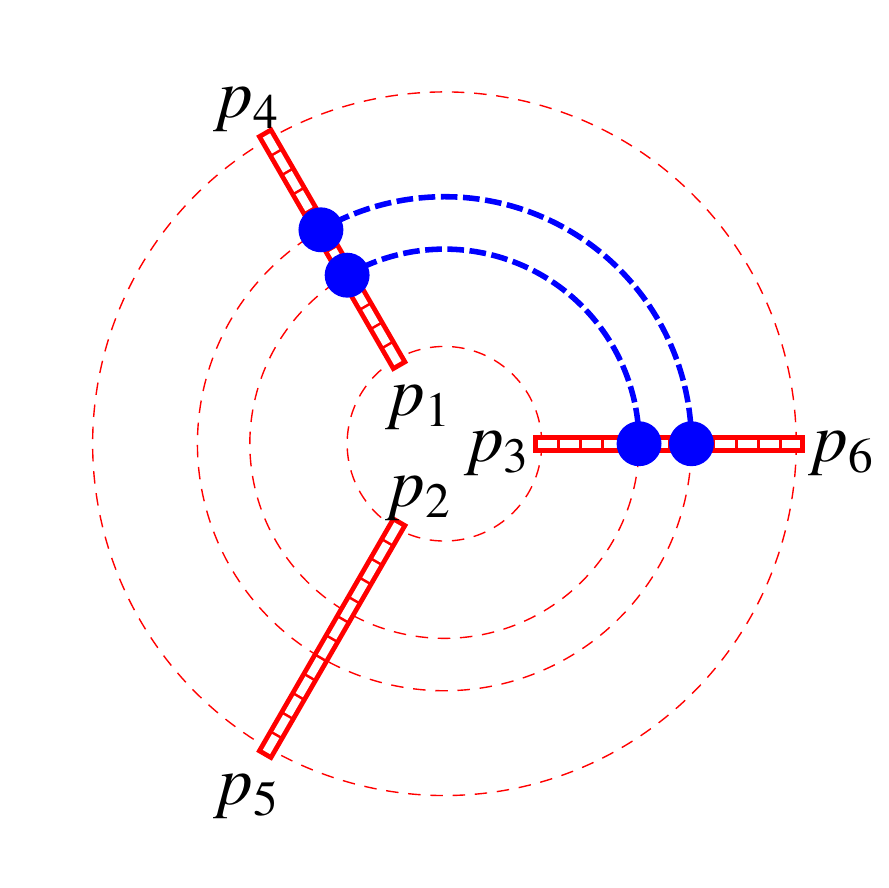}
\end{center}
\caption{First 13 effective diagrams contributing to the BKP gluon Green function. }
\label{13Terms}
\end{figure}

In this figure each blob at rapidity $y_i$ on a line with transverse momentum ${\bf p}_i$ represents a term of the form $e^{\omega ({\bf p}_i) (y_{i+1}-y_i)}$. For a line without any connections with other lines through the function $\xi$ we would have that the sum of all trajectories contributes with the term  $e^{\omega ({\bf p}_i) Y}$. Let us illustrate this with an example: with these effective Feynman diagrams we now find that the graphical representation of the first three iterations of the BKP equation corresponds to the one shown in the 13 terms of  Fig.~\ref{13Terms}.

\begin{figure}[h]
\begin{center}
\includegraphics[width=6cm]{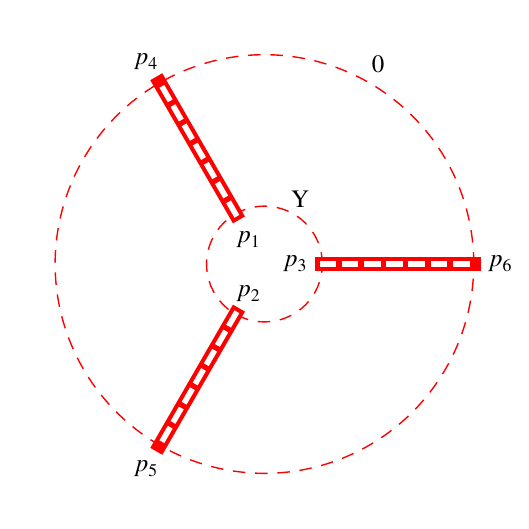}
\end{center}
\vspace{-1cm}
\begin{eqnarray}
&=&\delta^{(2)} \left({\bf p}_1-{\bf p}_4\right) \delta^{(2)} \left({\bf p}_2-{\bf p}_5\right) \delta^{(2)} \left({\bf p}_3-{\bf p}_6\right)  e^{(\omega({\bf p}_1) + \omega({\bf p}_2) + \omega({\bf p}_3)) Y} 
\nonumber
\end{eqnarray}
\caption{Contribution with 3 reggeized propagators and no Lipatov vertices $\xi$. }
\label{ThreeDeltas}
\end{figure}

\begin{figure}[h]
\begin{center}
\includegraphics[width=6cm]{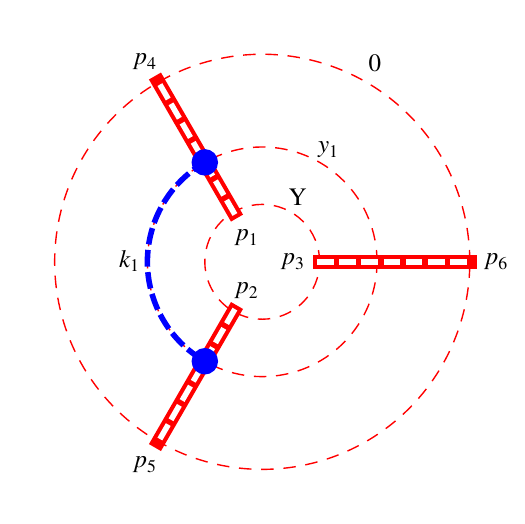}
\end{center}
\vspace{-1cm}
\begin{eqnarray}
&=& \int d^2 {\bf k}_1 \int_0^Y d y_1 \delta^{(2)} \left({\bf p}_3-{\bf p}_6\right) \delta^{(2)}
   \left({\bf k}_1+{\bf p}_1-{\bf p}_4\right) \delta^{(2)}
   \left(-{\bf k}_1+{\bf p}_2-{\bf p}_5\right)\nonumber\\
   &\times&  \xi
   \left({\bf p}_1,{\bf p}_2,{\bf p}_3,{\bf k}_1\right)  e^{\omega
   \left({\bf p}_3\right) Y} e^{
   \left(\omega \left({\bf k}_1+{\bf p}_1\right)+\omega
   \left({\bf p}_2-{\bf k}_1\right)\right)y_1}e^{\left(\omega
   \left({\bf p}_1\right)+\omega
   \left({\bf p}_2\right)\right)
   \left(Y-y_1\right)}  \nonumber
\end{eqnarray}
\caption{Effective Feynman diagram with one exchanged rung.}
\label{OneRung}
\end{figure}

\begin{figure}[h]
\begin{center}
\centering
\includegraphics[width=6cm]{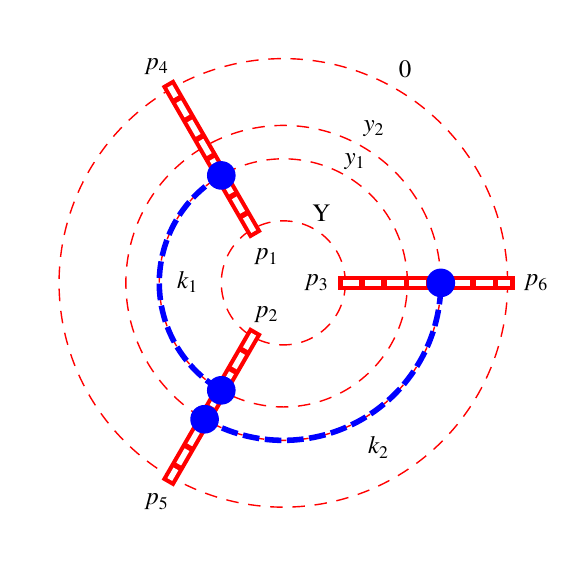}
\end{center}
\vspace{-1cm}
\begin{eqnarray}
&=&\int d^2 {\bf k}_1  \int d^2 {\bf k}_2 \int_0^Y d y_1 \int_0^{y_1} d y_2
\delta^{(2)} \left(k_1+p_1-p_4\right) \delta^{(2)}
   \left(-k_1+k_2+p_2-p_5\right) \nonumber\\
   &\times&  \delta^{(2)}
   \left(-k_2+p_3-p_6\right) \xi
   \left(p_1,p_2,p_3,k_1\right)  \xi
   \left(p_2-k_1,p_3,k_1+p_1,k_2\right) \nonumber\\
   &\times&   e^{ \left(\omega
   \left(p_1\right)+\omega
   \left(p_2\right)\right)
   \left(Y-y_1\right) }e^{ \omega
   \left(p_3\right)
   \left(Y-y_2\right) } e^{ \omega
   \left(k_1+p_1\right) y_1 }   e^{ \omega
   \left(p_2-k_1\right) \left(y_1-y_2\right) }  e^{ \left(\omega
   \left(-k_1+k_2+p_2\right)+\omega
   \left(p_3-k_2\right)\right)  y_2 }\nonumber
\end{eqnarray}
\caption{Effective Feynman diagram with two exchanged rungs.}
\label{TwoRungs}
\end{figure}

To better understand how to work with these effective Feynman diagrams, let us now discuss in more detail the structure of some of them. As we already pointed out, our initial condition, or first term of the iteration, corresponds to three delta functions, one per $t$-channel gluon propagator. Due to the action of those diagrams with Regge trajectories in the BKP kernel, this initial condition transforms into a first contribution to the BKP gluon Green function with modified gluon propagators, each of them picking up a Regge factor in the form as it is shown in Fig.~\ref{ThreeDeltas}. 
The set of momenta ${\bf p}_{i=1,2,3}$ are associated to a rapidity $Y$ and  ${\bf p}_{i=4,5,6}$  to a rapidity 0. This diagram corresponds to the first term in the expansion of Eq.~(\ref{IterativeEqn}). The structure of our effective Feynman diagrams in the high energy limit is richer when calculating the next terms in this equation. Let us consider now the case with one gluon rung as in the effective graph of Fig.~\ref{OneRung}.
We have to integrate over the phase space of the exchanged gluon with transverse momentum ${\bf k}_1$ and rapidity $y_1$, which can lie between 0 and the total rapidity $Y$. This integration is trivial due to the different delta functions present in the integrand. The integration over rapidity is a bit more complicated when we consider a diagram with more than one rung since then, in the integration limits, we should ensure rapidity ordering. This is shown in Fig.~\ref{TwoRungs} where we have to integrate over a two-gluon phase space. 

As we will see below, this way of rewriting the gluon Green function is very useful from a numerical point of view but each individual term with a fixed number of rungs in our sum is not infrared finite ($\lambda$ independent for small $\lambda$). This is clear since the terms with gluon Regge trajectories are exponentiated while the rungs are kept at a fixed order and the cancellation takes place order-by-order in the coupling. The full $\lambda$ independence is only achieved after we have summed over an infinite number of contributions. Fortunately this is not needed since, for a finite value of the coupling and $Y$, numerical convergence is achieved after summing up to a large but finite number of terms. This will be the subject of discussion in the coming Section.

\section{The Monte Carlo integration approach \& numerical results}

We will demonstrate the effectiveness of the approach described
before with a  concrete example. We want to evaluate Eq.~(\ref{IterativeEqn})
to obtain  $f \left({\bf p}_1,{\bf p}_2,{\bf p}_3, Y\right)$ for a given set
of ${\bf p}_1,{\bf p}_2,{\bf p}_3,{\bf p}_4,{\bf p}_5,{\bf p}_6$ and for different
rapidities $Y$. 
For concreteness, let us take the following two sets of values for
the different transverse 
momenta in the problem (all shown in polar coordinates such that
the first entry stands for the modulus of the momentum and the second
one for the azimuthal angle):
 \begin{align} 
& \,\, \, {\bf q} = (4, 0)                                                     &{\bf q} &= (31, 0) \nonumber \\
& {\bf p}_1 = (10, 0)                                               &{\bf p}_1  &= (10, 0) \nonumber \\
& {\bf p}_2  = (20, \pi)                                            &{\bf p}_2  &= (20, \pi) \nonumber \\
& {\bf p}_3 = ({\bf q}-{\bf p}_1)-{\bf p}_2 = (14,0)   &  {\bf p}_3 &= ({\bf q}-{\bf p}_1)-{\bf p}_2 = (41,0) \nonumber \\
& {\bf p}_4  = (20, 0)                                              &{\bf p}_4  &= (20, 0) \nonumber \\
& {\bf p}_5 =   (25, \pi)                                           &{\bf p}_5  &= (25, \pi) \nonumber \\
& {\bf p}_6 =({\bf q}-{\bf p}_4)-{\bf p}_5=(9,0)        &  {\bf p}_6 &=({\bf q}-{\bf p}_4)-{\bf p}_5=(36,0).
\label{momenta}
 \end{align}
 The  momentum transfer ${\bf q}$  in the LHS column has a relatively small value whereas in the RHS column it is much larger,
  ${\bf q} = (4, 0)$ and
  ${\bf q} = (31, 0)$ respectively. The moduli in the two-vectors 
  carry units of momentum, all of them being expressed in GeV. 
We remind the reader that 
\begin{align}
{\bf p}_1 + {\bf p}_2 + {\bf p}_3 = {\bf q} = {\bf p}_3 + {\bf p}_4 + {\bf p}_6\,.
\label{mom-equality}
\end{align}
 We will vary the rapidity $Y$ in the range $\left[1,\,\,5.5\right]$
  in steps of half a unit.

Once we have set the values for the momenta and the rapidity, we are ready to iterate 
and evaluate Eq.~(\ref{IterativeEqn}) for $n=1, 2, 3, ... , n_{\rm max}$, where $n_{\rm max}$ is a number after
which numerical convergence is reached or, in other words, 
$f \left({\bf p}_1,{\bf p}_2,{\bf p}_3, Y\right)$  does not increase any more. To be more precise,
for any $n>n_{\rm max}$ there will always be a possible further term to be considered but in reality its size 
gets smaller and smaller with increasing $n$ and their sum adds up to a tiny value that is much smaller than
the statistical uncertainty accumulated for adding the contributions up to $n_{\rm max}$.

After inspecting the momenta in Eq.~(\ref{momenta}), it is apparent that there is no contribution
to be added by considering only one iteration. Indeed, with only one rung connecting any
two reggeons, we cannot fulfil the Dirac delta functions as one can verify from Fig.~\ref{OneRung}.
Therefore, non-zero contributions to $f \left({\bf p}_1,{\bf p}_2,{\bf p}_3, Y\right)$ will appear
only after the second iteration once we consider two rungs. Moreover, any given diagram with
$i$ rungs will generate three new diagrams in the next iteration with $i+1$ rungs since there
are three pairs of reggeons we can connect with the new rung. This leads to a complete ternary 
tree structure. To be more specific, in Fig.~\ref{TwoRungs}, if we name the rungs
connecting ${\bf p}_1$ and ${\bf p}_2$ as  ``left" or L rungs, the ones  
connecting ${\bf p}_2$ and ${\bf p}_3$ as  ``middle" or M rungs and the ones
connecting ${\bf p}_1$ and ${\bf p}_3$ as  ``right" or R rungs we 
may represent the first 13 effective diagrams contributing to the BKP gluon
Green function after two iterations by the  following equivalent ternary tree
structure:\\
\vspace{1cm}\\
\begin{tikzpicture}[every node/.style={circle,draw},level 1/.style={sibling distance=40mm},level 2/.style={sibling distance=14mm}
]
\node {0}
child { 
node {L} 
child { node {LL}}
child { node {LM}}
child { node {LR}}
}
child { 
node {M} 
child {node {ML}}
child {node {MM}}
child {node {MR}}
}
child { 
node {R} 
child {node {RL}}
child {node {RM}}
child {node {RR}}
}
;
\label{tex-tree}
\end{tikzpicture}
\vspace{1cm}\\
\hspace{-.7cm} where each node of the tree is labeled using the L, M, R notation defined
above. In a similar fashion, we will name the individual effective diagrams using
the ordered L, M, R notation, {\it e.g.} ${\mathcal{D}}_{\text{M}}$, ${\mathcal{D}}_{\text{MRRLML}}$, etc.

For a given number of iterations $i$, it is obvious that we will have $3^i$ effective contributions. To be
specific,  for $n = 2, ..., 14$ rungs we have:
\begin{align}
&n \,\,\text{rungs} \hspace{2cm}& \text{number of diagrams}
\nonumber \\
&2&9 \nonumber \\
&3&27\nonumber \\
&4&81\nonumber \\
&5&243\nonumber \\
&6&729\nonumber \\
&7&2187\nonumber \\
&8&6561\nonumber \\
&9&19683\nonumber \\
&10&59049\nonumber \\
&11&177147\nonumber \\
&12&531441\nonumber \\
&13&1594323 \nonumber \\
&14&4782969
\label{nums}
\end{align}
All these\footnote{Nevertheless, there are diagrams that are identically zero, as we will show later on.} diagrams contribute to the gluon Green function. The diagrams 
with equal number of rungs $i$ have to be summed up and be integrated over the
momenta of the exchanged gluons ${\bf k}_j$ and the rapidities $y_j$ with $j=1,...,i$.
One can see in Eq.~(\ref{nums}) 
that considering 
$n_{\rm max} =14$ means that one is left
with an integrand containing more than 4.7 million diagrams
and which needs to be integrated over 14 momenta ${\bf k}_{1,...,14}$ and
14 rapidity values $y_{1,...,14}$. We perform the integrations using
a Monte Carlo computational approach for the needed number of rungs in each case.

\begin{figure}[h]
\vspace{-2cm}
\centering
\includegraphics[scale=.6]{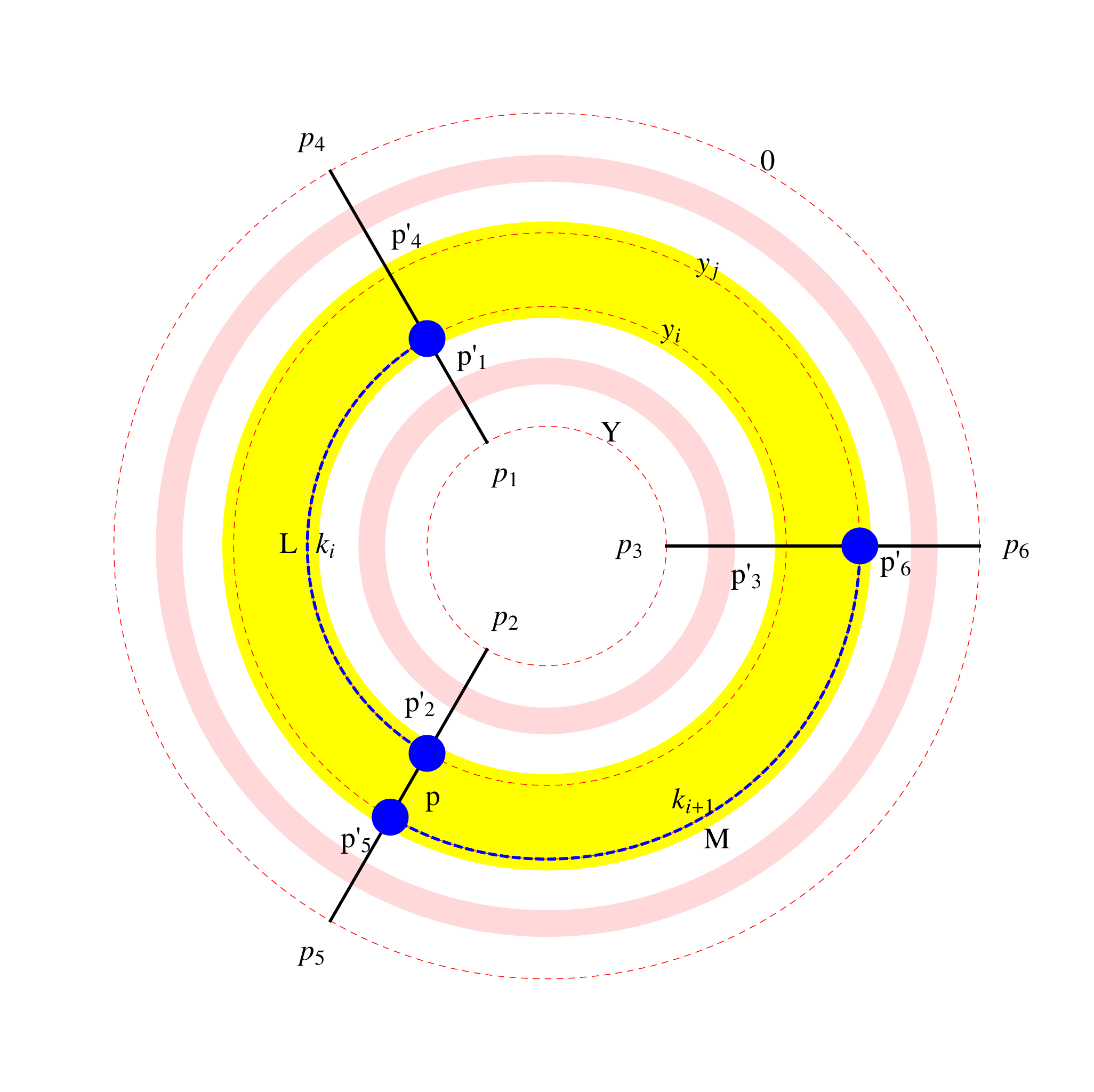}
\vspace{-1cm}
\caption{A LM rungs configuration that can ensure the fulfilment of the three Dirac delta functions
of the initial condition.
We will call such a configuration of two rungs in a BKP diagram a ``junction". There are in total
six different junctions: LM, ML, LR, RL, MR, RM.}
\label{fig:junction}
\end{figure}

It is noteworthy to indicate that two of the momenta integrations
are trivial if we remember that we have to fulfil the three Dirac delta functions
which still leaves us, in the general case, with $(n-2)$ two-dimensional non-trivial integrations regarding
the momenta.
To make this last point more transparent, let us consider Fig.~\ref{fig:junction} where the shaded regions in light red stand for an arbitrary number of exchanged gluons between the three reggeons.
With an slight abuse of notation, we will denote the momenta of the three reggeons
in the inner and outer above mentioned shaded regions by  ${\bf p}'_{1,2,3}$
and ${\bf p}'_{3,4,5}$ respectively.
Taking into account momentum conservation on the transverse plane in Fig.~\ref{fig:junction},
we have the following relations :
\begin{align}
{\bf p}'_4 &= {\bf p}'_1 + {\bf k}_i \nonumber \\
{\bf p}'_5 &= {\bf p}'_2 - {\bf k}_i + {\bf k}_{i+1} \nonumber \\
{\bf p}'_6 &= {\bf p}'_3 - {\bf k}_{i+1}
\end{align}
or equivalently
\begin{align}
{\bf k}_i &= {\bf p}'_4 - {\bf p}'_1 \nonumber \\
{\bf p}'_5 &= {\bf p}'_2 - {\bf k}_i + {\bf k}_{i+1} \nonumber \\
{\bf k}_{i+1} &= {\bf p}'_3 - {\bf p}'_6\,.
\end{align}
Substituting the values of ${\bf k}_i$ and ${\bf k}_{i+1}$ from the 
top and bottom into the middle equation
above, gives:
\begin{align}
{\bf p}'_5 &= {\bf p}'_2 - ({\bf p}'_4 - {\bf p}'_1) + ({\bf p}'_3 - {\bf p}'_6) \,\,\, \,\text{or}\nonumber \\
{\bf p}'_1 &+ {\bf p}'_2 + {\bf p}'_3 = {\bf p}'_4 + {\bf p}'_5 + {\bf p}'_6
\end{align}
which is trivially fulfilled since Eq.~(\ref{mom-equality}) holds.
It simply states that at any rapidity, if we add all the reggeon momenta,
we are bound to get the total momentum transfer ${\bf q}$ in the $t$-channel.
This is actually very important. It tells us that for any effective diagram that contributes 
to $f \left({\bf p}_1,{\bf p}_2,{\bf p}_3, Y\right)$,
we can allow the exchanged momenta to
take any random value apart from where we have what we call a ``junction" (highlighted by the yellow region in Fig.~\ref{fig:junction}):
a configuration of two subsequent rungs that can have any of the
LM, ML, LR, RL, MR, RM labels and which we denote respectively by
${\mathcal{J}}_{\text{LM}}$,
${\mathcal{J}}_{\text{ML}}$,
${\mathcal{J}}_{\text{LR}}$,
${\mathcal{J}}_{\text{RL}}$,
${\mathcal{J}}_{\text{MR}}$,
${\mathcal{J}}_{\text{RM}}$.
In other words, a junction is defined to be that
part of the effective diagram for which  the exchanged momenta
cannot take any random value but need to be set in a certain way
such that the global initial condition is fulfilled. 
In detail, the momenta of the exchanged gluons in the
six different junctions
are chosen as follows:

\begin{align}
&{\mathcal{J}}_{\text{LM}}:  &{\bf k}_i &= {\bf p}'_4 - {\bf p}'_1  \nonumber \\
&&{\bf k}_{i+1} &= {\bf p}'_3 - {\bf p}'_6\,  \nonumber \\
&&&  \nonumber \\
&{\mathcal{J}}_{\text{ML}}:  &{\bf k}_i &= {\bf p}'_3 - {\bf p}'_6  \nonumber \\
&&{\bf k}_{i+1} &= {\bf p}'_4 - {\bf p}'_1\,  \nonumber \\
&&&  \nonumber \\
&{\mathcal{J}}_{\text{LR}}:  &{\bf k}_i &= {\bf p}'_2 - {\bf p}'_5  \nonumber \\
&&{\bf k}_{i+1} &= {\bf p}'_3 - {\bf p}'_6\,  \nonumber \\
&&&  \nonumber \\
&{\mathcal{J}}_{\text{RL}}:  &{\bf k}_i &= {\bf p}'_3 - {\bf p}'_6  \nonumber \\
&&{\bf k}_{i+1} &= {\bf p}'_2 - {\bf p}'_5\,  \nonumber \\
&&&  \nonumber \\
&{\mathcal{J}}_{\text{MR}}:  &{\bf k}_i &= {\bf p}'_5 - {\bf p}'_2  \nonumber \\
&&{\bf k}_{i+1} &= {\bf p}'_4 - {\bf p}'_1\,  \nonumber \\
&&&  \nonumber \\
&{\mathcal{J}}_{\text{RM}}:  &{\bf k}_i &= {\bf p}'_4 - {\bf p}'_1  \nonumber \\
&&{\bf k}_{i+1} &= {\bf p}'_5 - {\bf p}'_2\,.
\label{eq:junctions}
\end{align}

Clearly, an effective diagram with high enough number of rungs may appear to 
have more than one
junction. In that case,
we consider as junction the first occurrence of one of the 
${\mathcal{J}}_{\text{LM}}$,
${\mathcal{J}}_{\text{ML}}$,
${\mathcal{J}}_{\text{LR}}$,
${\mathcal{J}}_{\text{RL}}$,
${\mathcal{J}}_{\text{MR}}$,
${\mathcal{J}}_{\text{RM}}$. For example, 
the diagram  ${\mathcal{D}}_\text{LLMRM}$ has only one junction, ${\mathcal{J}}_{\text{LM}}$, whereas the diagram
${\mathcal{D}}_\text{RRRRLLLR}$ has only the junction ${\mathcal{J}}_{\text{RL}}$.
Let us add here on a much more technical level that in order to find what the junction
is in a given diagram, we have chosen to evaluate the {\it adjacency list} for each diagram, which corresponds to a set of unordered lists which can be used to represent a finite graph. 
Defining the adjacency list will be of service for future works where we will be interested
in studying topologically different effective diagrams or to find ways to speed up the
computation time of the gluon Green function.
To conclude with, any diagram with no junction, that is,  any diagram ${\mathcal{J}}_{\text{Q}}$
 with Q being a sequence of only L or only M or only R is identically zero
 since the three Dirac delta functions of the initial condition cannot be fulfilled.
This is apparent if we note that all ${\bf p}_{1,2,3,4,5,6}$ we have chosen are different from each other. 

\begin{figure}[h]
\begin{center}
\includegraphics[width=12cm]{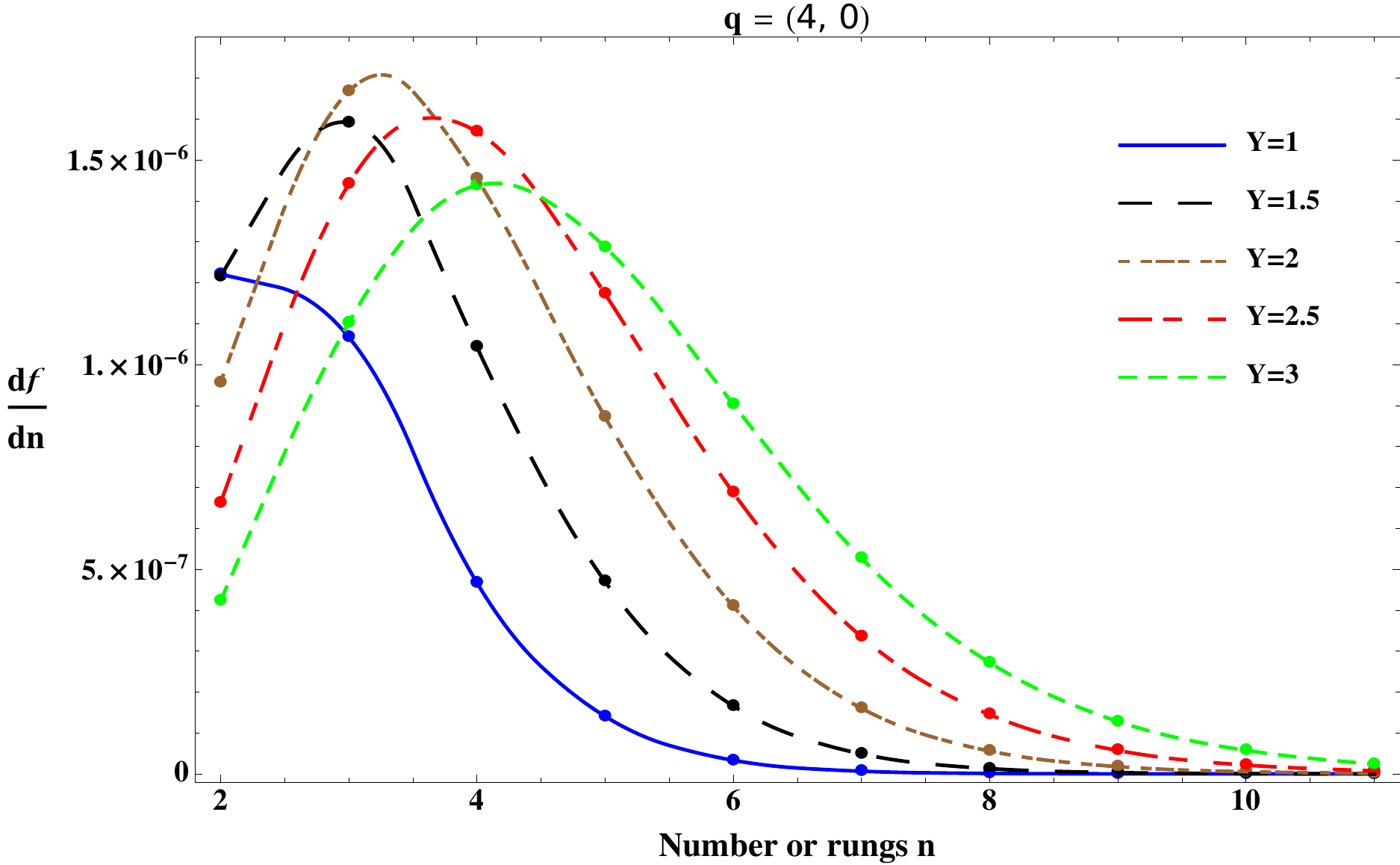}
\caption{``Multiplicity" plot for $ q=4$ GeV and smaller values of rapidity Y.}
\label{fig:multi1}
\end{center}
\end{figure}

\begin{figure}[h]
\begin{center}
\includegraphics[width=12cm]{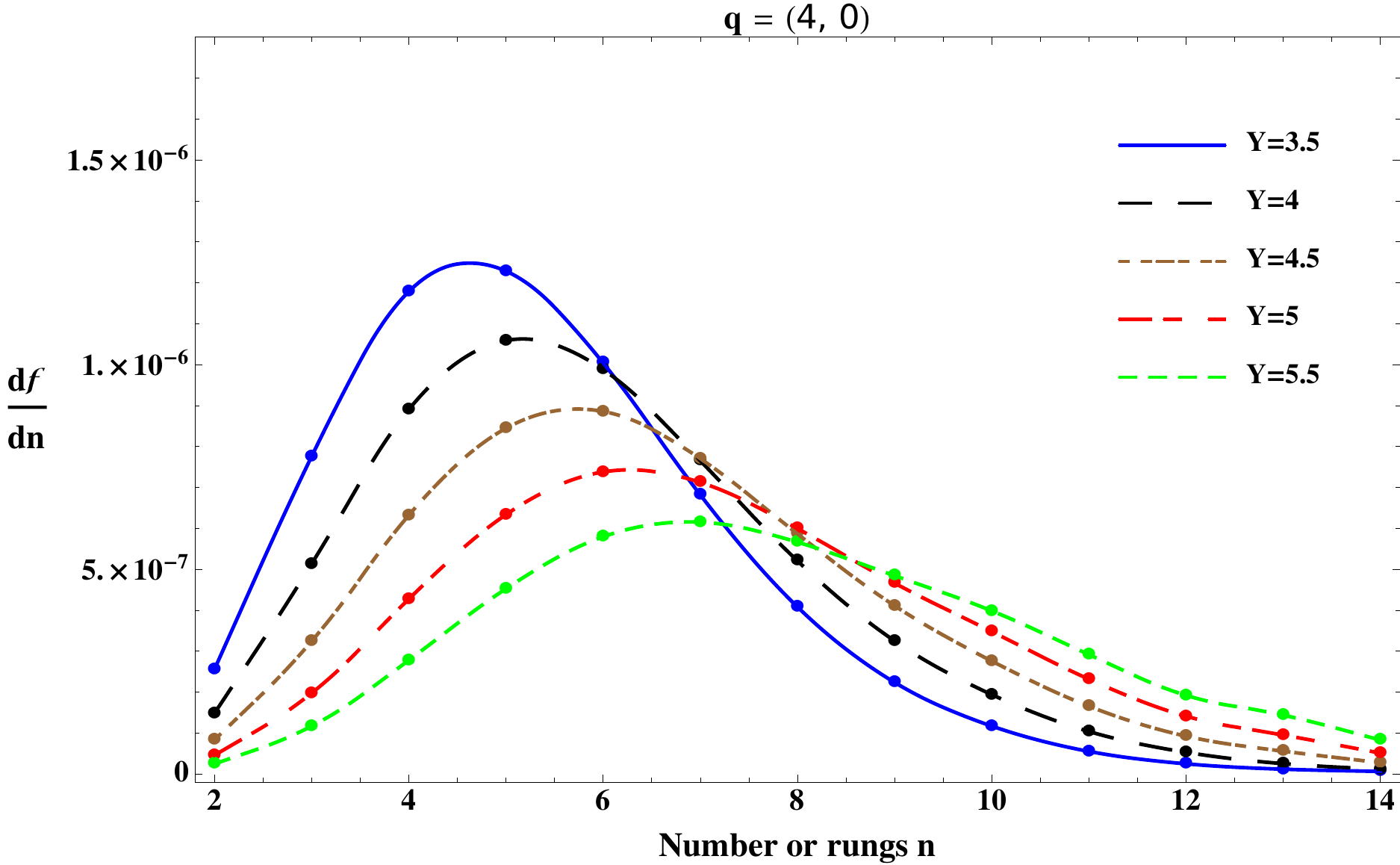}
\caption{``Multiplicity" plot for $ q=4$ GeV and larger values of rapidity Y.}
\label{fig:multi2}
\end{center}
\end{figure}

\begin{figure}[h]
\begin{center}
\includegraphics[width=12cm]{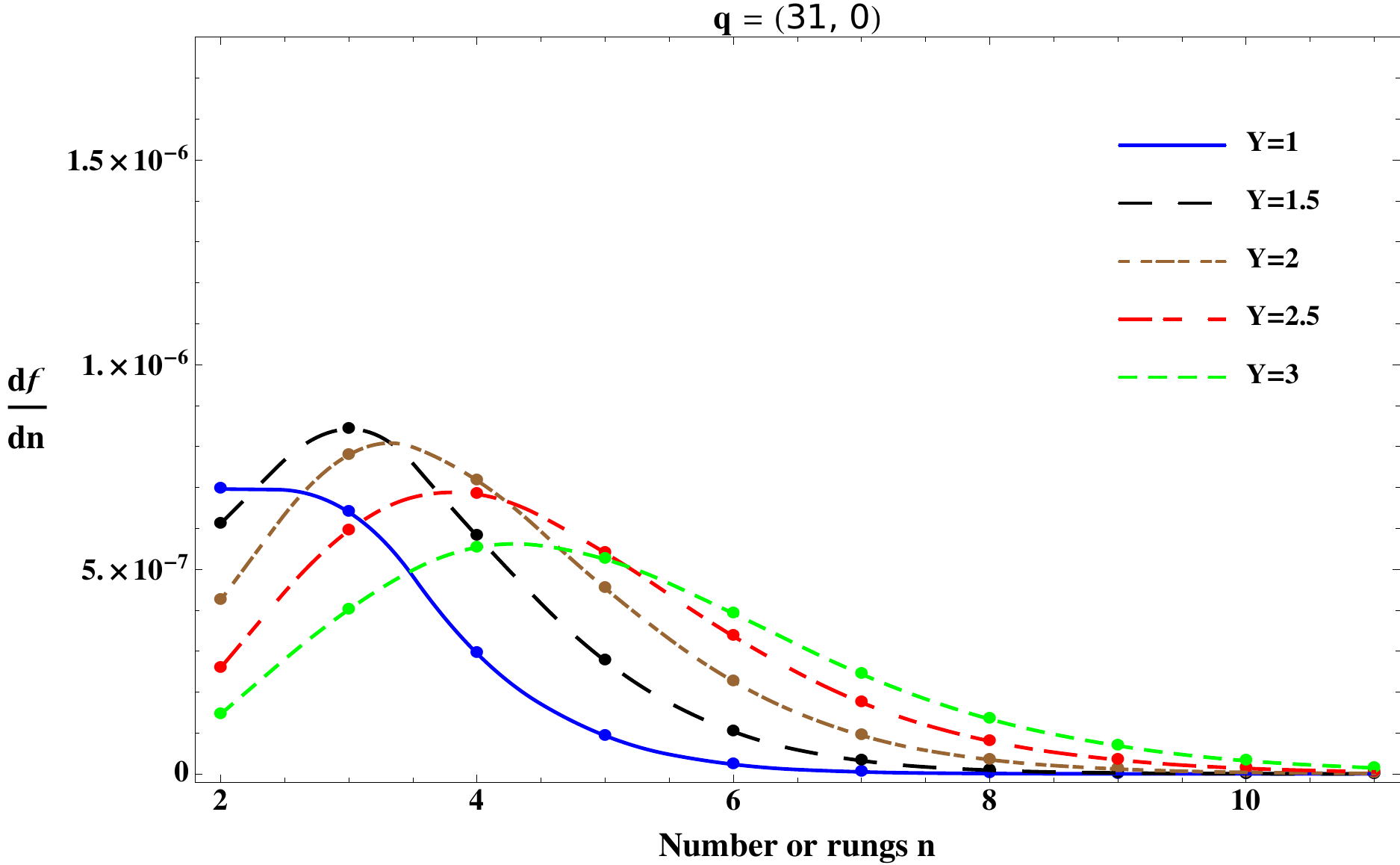}
\caption{``Multiplicity" plot for $ q=31$ GeV and smaller values of rapidity Y.}
\label{fig:large-multi1}
\end{center}
\end{figure}

\begin{figure}[h]
\begin{center}
\includegraphics[width=12cm]{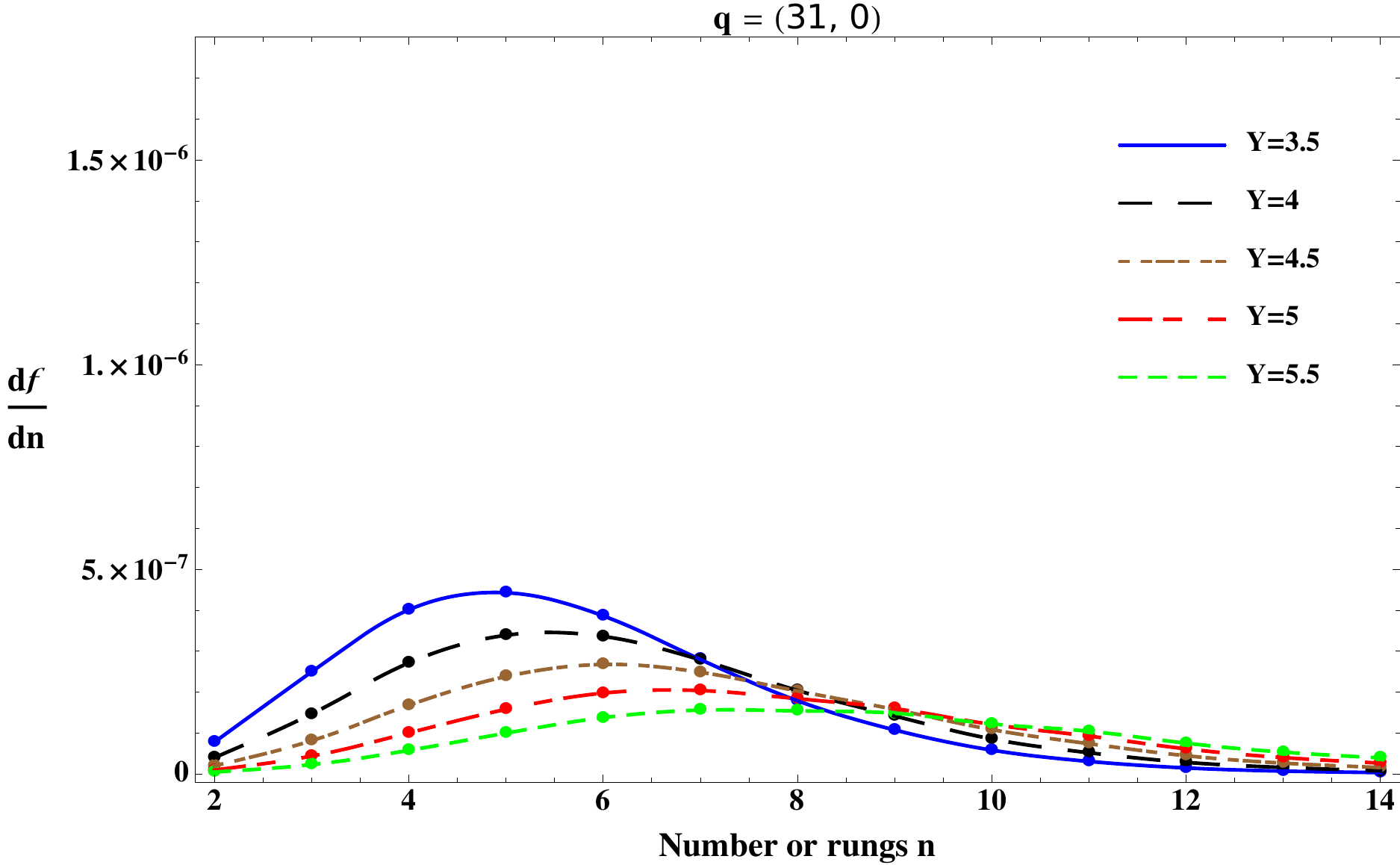}
\caption{``Multiplicity" plot for $ q=31$ GeV and larger values of rapidity Y.}
\label{fig:large-multi2}
\end{center}
\end{figure}

Having the machinery described above at hand, we were able to proceed to the
numerical integration of all the individual contributions in order to find numerical convergence.
We went up to 14 rungs although for most of the smaller rapidity values in the range 
$\left[1,\,\,5.5\right]$ 11 rungs were enough.
The first thing we noticed is that when we plot the contributions from each
iteration for a given $Y$ 
(what we will loosely call ``multiplicity" plot), they seem to follow a Poisson-like
distribution. At some $n$ they have a global maximum and they fall as $n$ increases.
To make this feature more apparent, we present here two figures for each
value of the total momentum transfer in the two configurations chosen to illustrate our results in Eq.~(\ref{momenta}), that is for ${\bf q} = (4, 0)$ (Figs.~\ref{fig:multi1},~\ref{fig:multi2})
and for ${\bf q} = (31, 0)$ (Figs.~\ref{fig:large-multi1},~\ref{fig:large-multi2}).
Figs.~\ref{fig:multi1},~\ref{fig:large-multi1} are for
$Y$ in the range $\left[1,\,\,3\right]$ and 
Figs.~\ref{fig:multi2},~\ref{fig:large-multi2} are for $Y$
taking values in $\left[3.5,\,\,5.5\right]$. In all four ``multiplicity" plots, we show
the data points along with an interpolation to make more visible the Poisson-like
features of the distributions. In all these plots, we kept the vertical axis range fixed to make comparisons easier.

From a first inspection of the four ``multiplicity" plots, we conclude that the gluon Green function
is noticeably smaller when the total momentum transfer is larger, ${\bf q} = (31, 0)$.
This is connected to the fact that while ${\bf p}_{1,2}$ and ${\bf p}_{4,5}$ are unchanged
when we change ${\bf q}$, the momenta ${\bf p}_{3}$ and ${\bf p}_{6}$ do change and actually
they get considerably larger, see Eq.~(\ref{momenta}).
The peak of the distribution moves similarly to larger values of $n$ for both values of ${\bf q}$ as we 
increase $Y$. Moreover, the peak gets lower as the energy rises and the distributions
get much broader while for ${\bf q} = (31, 0)$ the peak is delayed (occurs at larger $n$)
with respect to what happens for ${\bf q} = (4, 0)$.

\begin{figure}[h]
\begin{center}
\includegraphics[width=12cm]{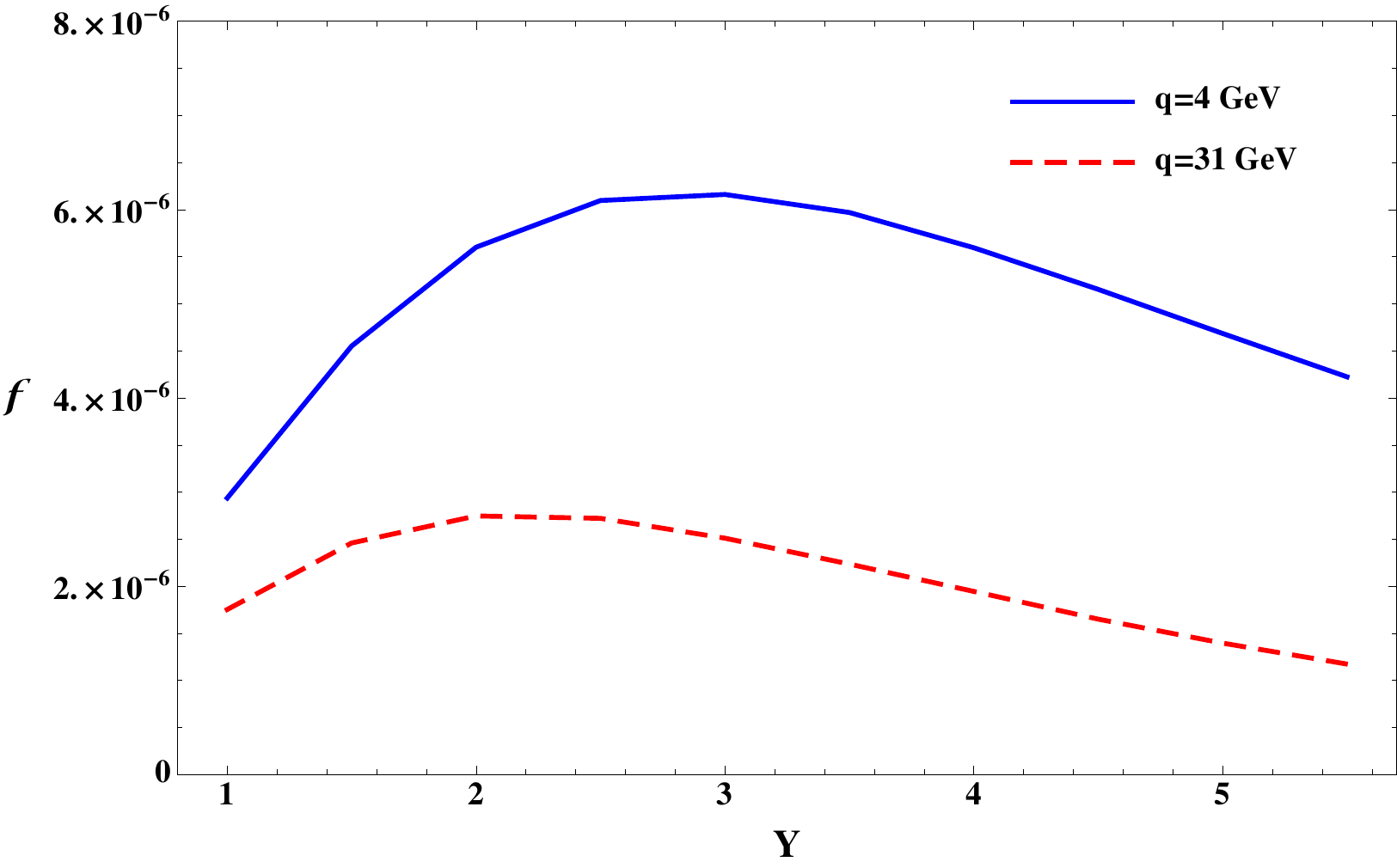}
\caption{Energy dependence of $f$ for $q=4$ GeV (blue continuous line) and 
$q=31$ GeV (red dashed line).}
\label{fig:energy}
\end{center}
\end{figure}

It is very interesting to plot the rapidity dependence of the gluon Green function
for the two values of ${\bf q}$, see Fig.~\ref{fig:energy}. What we observe is that
both curves initially increase with $Y$, they reach a maximum and then they
start falling. The maximum occurs later in $Y$ for ${\bf q} = (4, 0)$.
Since we are well below the asymptotic region, we cannot say much about 
the large $Y$ behaviour of the gluon Green function besides the fact that it
reaches a maximum value at a certain rapidity and then it drops monotonically. 
When compared to the BFKL pomeron Green function~\cite{BFKL}, which grows rapidly with $Y$, in the odderon case we have a kernel where the $\xi$ function in Eq.~(\ref{xifunction}) has an extra factor of 2 in the denominator. This is because we operate, in each iteration of the kernel,  with the adjoint representation. In this representation the contributions from the Regge gluon trajectories, which tend to lower the value of cross sections or the Green function in this case, are then enhanced with respect to the ``rung" contributions, which typically ``push upwards" in $Y$ the solution of the corresponding integral equation.  This is the mechanism driving the above mentioned decrease with $Y$ of our odderon solution. 
This feature of the BKP kernel is of great help when applying our Monte Carlo integration techniques since it allows to reach convergence with fewer number of iterations of the kernel  than in the BFKL case, which carries a singlet representation in its kernel (we have investigated this point making use of the Monte Carlo event generator {\tt BFKLex}~\cite{BFKLex}).  

\section{Conclusions \& outlook}

A novel method of solution of the Bartels-Kwiecinski-Praszalowicz (BKP) equation has been 
presented. This approach relies on the numerical integration of iterated integrals in transverse momentum and rapidity space. We have applied it to the BKP equation with three reggeized gluons in the $t$-channel, the so-called odderon case. We have shown that numerical convergence of the solution is achieved after applying the BKP ternary kernel on the initial condition, corresponding to three off-shell gluon propagators, a finite number of times for a given value of the strong coupling and the total center-of-mass energy encoded in the rapidity variable $Y$.  We have shown that the gluon Green function for reggeized gluons grows with $Y$ for small values of this variable to then rapidly decrease at higher $Y$. This is compatible with previous approaches where the odderon intercept has been argued to be of ${\cal O} (1)$~\cite{Bartels:1999yt}. This stems from the competition between terms holding the gluon Regge trajectory and those related to iterations of the square of the so-called Lipatov's vertex, which in the adjoint representation, characteristic of the BKP kernel, is won by the former. We have performed some explicit calculations of the odderon gluon Green function for different values of the momentum transfer ${\bf q}$  finding a qualitatively very similar behavior at small and large values of ${\bf q}$ in terms of growth with $Y$ and distribution in the number of needed iterations to reach numerical convergence. The main difference lies in having smaller values for the Green function in the latter case.

The formalism here described can be applied to the BKP equation with a higher number of exchanged reggeons. It can also be used beyond the leading logarithmic approximation~\cite{Bartels:2012sw} and for cases with a total $t$-channel color projection not being in the singlet but in the adjoint representation. This is very important for the calculation of scattering amplitudes in $N=4$ supersymmetric theories in the Regge limit~\cite{Bartels:2008ce}. Our approach also has obvious applications in the study of phenomenological cross sections devoted to the search of the elusive odderon at hadron colliders~\cite{N.Cartiglia:2015gve}. All of these lines of investigation are part of our future plans for applications of this work.

\vspace{1cm}
\begin{flushleft}
{\bf \large Acknowledgements}
\end{flushleft}
GC acknowledges support from the MICINN, Spain, 
under contract FPA2013-44773-P, ASV thanks the Spanish Government 
(MICINN (FPA2015-65480-P)) and both to the Spanish MINECO Centro de Excelencia Severo Ochoa Programme (SEV-2012-0249) for support.

\end{document}